# New Lower Bounds on the Self-Avoiding-Walk Connective Constant

Takashi Hara
Department of Applied Physics
Tokyo Institute of Technology
Oh-Okayama, Meguro-ku, Tokyo 152
Japan
E-mail address: `HARA@APPANA.AP.TITECH.AC.JP`

Gordon Slade
Department of Mathematics and Statistics
McMaster University
Hamilton, Ontario
Canada L8S 4K1
E-mail address: `SLADE@AURORA.MATH.MCMASTER.CA`

Alan D. Sokal
Department of Physics
New York University
4 Washington Place
New York, NY 10003
USA
E-mail address: `SOKAL@ACF3.NYU.EDU`

February 11, 1993

**Running head:** Lower Bounds on the SAW Connective Constant

## Abstract

We give an elementary new method for obtaining rigorous lower bounds on the connective constant for self-avoiding walks on the hypercubic lattice $\mathbb{Z}^d$. The method is based on loop erasure and restoration, and does not require exact enumeration data. Our bounds are best for high $d$, and in fact agree with the first four terms of the $1/d$ expansion for the connective constant. The bounds are the best to date for dimensions $d \geq 3$, but do not produce good results in two dimensions. For $d = 3, 4, 5, 6$, respectively, our lower bound is within $2.4\%$, $0.43\%$, $0.12\%$, $0.044\%$ of the value estimated by series extrapolation.

# 1   Introduction

An $n$-step self-avoiding walk on the hypercubic lattice $\mathbb{Z}^d$ is a sequence $\omega = (\omega(0), \omega(1), \ldots, \omega(n))$ of points in $\mathbb{Z}^d$, with $\omega(i)$ and $\omega(i+1)$ separated by Euclidean distance one, subject to the constraint that $\omega(i) \neq \omega(j)$ for $i \neq j$. Unless otherwise stated we take $\omega(0) = 0$. The self-avoiding walk provides a model of a polymer molecule with excluded volume. Also, its equivalence to the $N = 0$ limit of the $N$-vector model has made it an important test case in the theory of critical phenomena.

Let $c_n$ denote the number of $n$-step self-avoiding walks. It has been known for almost forty years [1] that the limit $\mu = \lim_{n \to \infty} c_n^{1/n}$ exists and is finite and positive. Moreover the subadditivity argument showing existence of this limit also shows that $\mu = \inf_{n \geq 1} c_n^{1/n}$. This limit is known as the *connective constant*, and is the analogue of a critical temperature for the $N$-vector model. Roughly speaking, $\mu$ measures the number of sites available for the next step of a long self-avoiding walk. The connective constant is admittedly of lesser interest than the critical exponents, because it is lattice-dependent while the critical exponents are universal. Nevertheless, much work has been done in finding rigorous upper and lower bounds for the connective constant, principally on two- and three-dimensional lattices. A review of work through 1982 is given by Guttmann [2].

Recently, two of us [3, 4] proved that the critical exponents for self-avoiding walks in dimensions $d \geq 5$ take their mean-field values ($\gamma = 1$, $\nu = 1/2$, $\eta = 0$). One key ingredient in this proof was an accurate numerical lower bound on the connective constant $\mu$. Unfortunately, we were unable to prove such a numerical bound with existing methods; in fact the previous methods give estimates which deteriorate as $d \to \infty$ (see Section 6). Therefore, we were led to develop a new method for obtaining lower bounds on $\mu$, using loop erasure and restoration. This method (with the improvements presented here) provides bounds that agree with the first four terms of the $1/d$ expansion; for $d \geq 4$ they are within $0.43\%$ of the best numerical estimate of $\mu$, and even for $d = 3$ they are within $2.39\%$ (greatly improving the best previous lower bounds). We therefore thought it worthwhile to develop these methods in detail; that is the goal of the present paper. The method presented here involves a conceptual simplification of the methods used in [4], and also leads to better lower bounds. Remarkably, even the most elementary of our new methods leads to a better lower bound in $d = 3$ than has been obtained previously, including the enumeration bound 4.352 of [2]. Table 1 summarizes our best bounds and compares them to previously obtained bounds on $\mu$.

| $d$ | previous bound | this work | estimate | upper bound |
|---|---|---|---|---|
| 2 | $2.620\,02^a$ | $2.305\,766$ | $2.638\,158\,5\,(10)^d$ | $2.695\,76^b$ |
| 3 | $4.437\,33^c$ | $4.572\,140$ | $4.683\,907\,(22)^e$ | $4.756^b$ |
| 4 | $6.718\,00^c$ | $6.742\,945$ | $6.7720\,(5)^f$ | $6.832^b$ |
| 5 | $8.821\,28^c$ | $8.828\,529$ | $8.8386\,(8)^g$ | $8.881^b$ |
| 6 | $10.871\,199^c$ | $10.874\,038$ | $10.8788\,(9)^g$ | $10.903^b$ |

Table 1: Rigorous lower and upper bounds on the hypercubic-lattice connective constant $\mu$, together with estimates of actual values, for dimensions 2,3,4,5,6. Errors in the last digit(s) are shown in parentheses.
a) Conway and Guttmann [5], b) Alm [6], c) Hara and Slade [4], d) Guttmann and Enting [7, 8], e) Guttmann [9], f) Guttmann [10], g) Guttmann [11].

The evaluation of our bounds requires some numerical computation, for which we have obtained rigorous error estimates. All numerical values reported in this paper are accurate up to rounding of the last digit, except for lower bounds on $\mu$ which have been truncated so as to provide true lower bounds.

Our methods can be applied to SAWs on any regular lattice. But for simplicity we restrict attention here to the hypercubic lattice $\mathbb{Z}^d$.

The plan of this paper is as follows: In Section 2 we describe the method of loop erasure and restoration, and systematize the lower bounds on $\mu$ that can be obtained from it. These bounds involve generating functions of random walks with taboo sets, and in Section 3 we show how these taboo generating functions can be computed in terms of the massless-free-field lattice propagator. In Section 4 we discuss some aspects of the lower bounds on $\mu$ obtained. In Section 5 we remark on a different method for proving lower bounds on $\mu$, based on comparison with the Ising model. In Section 6 we show that our best bounds agree with the $1/d$ expansion for $\mu$ through order $d^{-2}$. In Appendix A we summarize our methods for the rigorous numerical calculation of quantities involving the free-field lattice propagator, and give $1/d$ expansions for various simple-random-walk quantities.

In a separate paper [12], two of us give a rigorous $1/d$ expansion for $\mu$ through order $d^{-3}$, along with a similar expansion for the critical point of nearest-neighbour Bernoulli bond percolation.

# 2 Loop erasure and restoration

## 2.1 Definitions

To describe our loop-erasure-and-restoration method, we need to introduce a number of generating functions. For this we need several definitions.

An *n-step walk* ($n \geq 0$) is an ordered sequence $\omega = (\omega(0), \ldots, \omega(n))$ of points in $\mathbb{Z}^d$ such that each point is a nearest neighbour of its predecessor, i.e. $|\omega(i) - \omega(i-1)| = 1$ for $1 \leq i \leq n$. We denote the number of steps in a walk $\omega$ by $|\omega|$. The walk $\omega$ is said to be a *memory-$\tau$ walk*, if $\omega(i) \neq \omega(j)$ for all $i, j$ satisfying $0 < |i-j| \leq \tau$. We denote by $\Omega_\tau(x, y)$ the union over all $n = 0, 1, 2, \ldots$ of the set of memory-$\tau$ $n$-step walks from $\omega(0) = x$ to $\omega(n) = y$. Thus, $\Omega_0(x, y)$ is the set of all walks from $x$ to $y$, $\Omega_2(x, y)$ is the set of all walks having no immediate reversals, $\Omega_4(x, y)$ is the set of all walks having neither immediate reversals nor elementary squares, and so on. The elements of $\Omega_0$ are called *simple* (or *ordinary*) *random walks*. We denote by $\Omega(x, y) \equiv \cap_{n=0}^\infty \Omega_n(x, y)$ the set of all self-avoiding walks (of any number of steps) which begin at $x$ and end at $y$. Finally, we denote by $\Omega(x, \bullet) \equiv \cup_{y \in \mathbb{Z}^d} \Omega(x, y)$ the set of all self-avoiding walks (of any number of steps) which begin at $x$ and end anywhere.

A walk $\omega = (\omega(0), \ldots, \omega(n))$ is said to be a *loop* if $\omega(0) = \omega(n)$. We write $\mathcal{L}_\tau(x) = \Omega_\tau(x, x)$ for the set of memory-$\tau$ loops starting and ending at $x$. Note that such loops are allowed to pass through $x$ many times, and that $\mathcal{L}_\tau(x)$ includes the zero-step walk $\omega = (x)$. Note also that the memory-$\tau$ constraint does *not* apply mod $n$: for example, $\omega(n-1)$ *is* permitted to equal $\omega(1)$.

Given a nonnegative real number $\beta$, we define the *generating function* (or *two-point function* or *Green function*) for memory-$\tau$ walks,

$$C_\tau(x, y; \beta) = \sum_{\omega \in \Omega_\tau(x,y)} \beta^{|\omega|}, \tag{2.1}$$

and for self-avoiding walks,

$$G(x, y; \beta) = \sum_{\omega \in \Omega(x,y)} \beta^{|\omega|}. \tag{2.2}$$

Denoting the number of $n$-step memory-$\tau$ walks (starting at the origin and ending anywhere) by $c_{n,\tau}$, we also define the susceptibilities

$$\chi_\tau(\beta) \equiv \sum_{x \in \mathbb{Z}^d} C_\tau(0, x; \beta) = \sum_{n=0}^\infty c_{n,\tau} \beta^n \tag{2.3}$$

and

$$\chi(\beta) \equiv \sum_{x \in \mathbb{Z}^d} G(0, x; \beta) = \sum_{n=0}^\infty c_n \beta^n. \tag{2.4}$$

For $\beta \geq 0$ the sums (2.1)–(2.4) are always well-defined, although they will be $+\infty$ for sufficiently large $\beta$. In fact, since $\lim_{n \to \infty} c_n^{1/n} = \mu$ and $c_n \geq \mu^n$, we have

$$(1 - \beta\mu)^{-1} \leq \chi(\beta) < \infty \quad \text{for } 0 \leq \beta < \mu^{-1} \tag{2.5a}$$

$$\chi(\beta) = \infty \quad \text{for } \beta \geq \mu^{-1}. \tag{2.5b}$$

Similarly, the same subadditivity argument which implies existence of the limit defining $\mu$ can also be used to prove that

$$\mu_\tau \equiv \lim_{n \to \infty} c_{n,\tau}^{1/n} = \inf_{n \geq 1} c_{n,\tau}^{1/n} \qquad (2.6)$$

for all $0 \leq \tau < \infty$. It follows from (2.6) that

$$(1 - \beta\mu_\tau)^{-1} \leq \chi_\tau(\beta) < \infty \qquad \text{for } 0 \leq \beta < \mu_\tau^{-1} \qquad (2.7a)$$
$$\chi_\tau(\beta) = \infty \qquad \text{for } \beta \geq \mu_\tau^{-1}. \qquad (2.7b)$$

Clearly $\mu_0 \geq \mu_2 \geq \mu_4 \geq \ldots \geq \mu$. A subadditivity argument can be used to show that $\lim_{\tau \to \infty} \mu_\tau = \mu$ (see for example Lemma 1.2.3 of [13]). Since $c_{n,0} = (2d)^n$ and $c_{n,2} = 2d(2d-1)^{n-1}$, we have $\mu_0 = 2d$ and $\mu_2 = 2d - 1$. The value of $\mu_4$ is shown in [14] to be given by the unique positive root of the equation

$$x^3 - 2(d-1)x^2 - 2(d-1)x - 1 = 0. \qquad (2.8)$$

Although methods are described in [14] by which in principle $\mu_\tau$ can be computed for $\tau \geq 6$, in practice these methods are difficult to carry out in general dimensions.

To compute our lower bounds on $\mu$ we will need the numerical values of $C_\tau(0, x; \mu_\tau^{-1})$ for a finite collection of sites $x$. For $\tau = 0$ this is given by the well-known Fourier integral ("free-field lattice propagator")

$$C_0(0, x; \beta) = \int_{[-\pi,\pi]^d} \frac{d^d k}{(2\pi)^d} \frac{e^{ik \cdot x}}{1 - 2d\beta\widehat{D}(k)}, \qquad (2.9)$$

where

$$\widehat{D}(k) \equiv \frac{1}{d} \sum_{i=1}^{d} \cos k_i, \qquad k = (k_1, \ldots, k_d). \qquad (2.10)$$

This expression is valid for $0 \leq \beta \leq \frac{1}{2d}$. At the critical point $\beta = \mu_0^{-1} = \frac{1}{2d}$, the integral (2.9) is finite for $d > 2$ but infinite for $d \leq 2$. An effective means for computing the numerical values of (2.9) to high precision for $d > 2$ is discussed in Appendix A.

To study $d \leq 2$ (of course it is $d = 2$ which is of interest here), we introduce the *potential kernel* [15, 16]

$$\Delta_0(x; \beta) \equiv C_0(0, 0; \beta) - C_0(0, x; \beta), \qquad (2.11)$$

which remains finite in all dimensions, as $\beta \uparrow \frac{1}{2d}$. Indeed, it is an immediate consequence of the Lebesgue dominated convergence theorem that $\Delta_0(x) \equiv \lim_{\beta \uparrow \frac{1}{2d}} \Delta_0(x; \beta)$ is given by the absolutely convergent Fourier integral

$$\Delta_0(x) = \int_{[-\pi,\pi]^d} \frac{d^d k}{(2\pi)^d} \frac{1 - \cos(k \cdot x)}{1 - \widehat{D}(k)}. \qquad (2.12)$$

In dimension $d = 1$, an easy calculation yields $\Delta_0(x) = |x|$. Remarkably, in dimension $d = 2$ this integral can also be performed analytically for any $x$ [15, Section 15]: for example, $\Delta_0(e_1) = 1$, $\Delta_0(2e_1) = 4 - 8/\pi$ and $\Delta_0(e_1 + e_2) = 4/\pi$. [Here $e_1, e_2$ are the canonical unit vectors in $\mathbb{Z}^2$.]

For $\tau = 2$ the two-point function can be evaluated using the identity

$$C_2(0, x; \beta) = \frac{1 - \beta^2}{1 + (2d - 1)\beta^2} C_0(0, x; \beta/[1 + (2d - 1)\beta^2]) , \qquad (2.13)$$

which was derived using convolution methods in [17]. At the critical point $\beta = \mu_2^{-1} = \frac{1}{2d-1}$, this reduces to

$$C_2(0, x; \tfrac{1}{2d-1}) = \frac{2d - 2}{2d - 1} C_0(0, x; \tfrac{1}{2d}) . \qquad (2.14)$$

Unfortunately we do not know how to compute the numerical values of $C_\tau(0, x; \mu_\tau^{-1})$ for $\tau \geq 4$; because of this we primarily restrict attention in what follows to $\tau = 0, 2$.

We remark that unlike the finite-memory case, it is believed that the self-avoiding-walk critical two-point function $G(0, x; \mu^{-1})$ is finite for all $x$ in all dimensions, including $d = 1, 2$. This has been proven for $d \geq 5$ in [3, 4] — and of course it is trivial in $d = 1$ — but it remains unproven in dimensions 2, 3, and 4. It has been known for some time that $G(0, x; \beta) = \infty$ for $\beta > \mu^{-1}$ [18].

## 2.2 Identities

We would like now to establish an inequality relating the two-point functions (2.1) and (2.2). For this, we recall the following loop-erasure algorithm, which has been studied in detail by Lawler [16]. Given a walk $\omega \in \Omega_\tau(x, y)$, we can associate to it a (typically shorter) self-avoiding walk $\rho \in \Omega(x, y)$ by erasing loops in an appropriate sequence. We begin by finding the *last* time $t_1$ such that $\omega(t_1) = \omega(0) = x$, and then erase the sites $\omega(1), \omega(2), \ldots, \omega(t_1)$ from $\omega$, producing a walk which we call $\rho^{(1)}$. In other words we have erased the largest possible loop at the site $x$, namely $L_0 = (\omega(0), \ldots, \omega(t_1))$. [If $\omega$ does not visit $x$ more than once, then we can think of having erased a trivial loop.] The walk $\rho^{(1)}$ does not visit $x$ more than once, but it may visit the site $\rho^{(1)}(1)$ repeatedly. Let $t_2$ denote the *last* time that $\rho^{(1)}(t_2) = \rho^{(1)}(1)$, and erase the sites $\rho^{(1)}(2), \ldots, \rho^{(1)}(t_2)$ as before. Note that the erased loop $L_1 = (\rho^{(1)}(1), \ldots, \rho^{(1)}(t_2))$ cannot pass through $\omega(0) = x$. This procedure gives rise to a walk $\rho^{(2)}$, which does not visit $\rho^{(2)}(0)$ or $\rho^{(2)}(1)$ more than once. We repeat this procedure successively for $\rho^{(2)}$, $\rho^{(3)}$, etc., until arriving at a result which is devoid of loops, or in other words which is self-avoiding. For each $\tau$, this defines a one-to-one mapping from $\Omega_\tau(x, y)$ into the set $\mathcal{R}_\tau(x, y)$ whose elements are of the form $(\rho, L_0, L_1, \ldots, L_n)$, where $\rho \in \Omega(x, y)$ is an $n$-step self-avoiding walk (for some $n$) and each $L_i \in \mathcal{L}_\tau(\rho(i))$. We refer to $\rho$ as the self-avoiding *backbone* of $\omega$.

In fact it is not difficult to see precisely what the image of this mapping is, or in other words to see exactly which elements of $\mathcal{R}_\tau(x, y)$ can be produced by this procedure, at least for $\tau = 0$ or 2. One way to do so is to try to reverse the procedure, by beginning with an element of $\mathcal{R}_\tau(x, y)$ and associating to it the walk $\omega$ which is given by first following the steps of $L_0$, then taking the first step of $\rho$, then taking the steps of $L_1$, then taking the second step of $\rho$, and so on. For $\tau = 0$, each possible simple random walk from $x$ to $y$ can be obtained in precisely one way by this procedure, provided that the loop $L_i$ attached at $\rho(i)$ does not intersect any of the previous sites $\rho(0), \ldots, \rho(i-1)$, for all $i = 1, \ldots, |\rho|$. For $\tau = 2$ the situation is similar but slightly more involved: the attached loops must again avoid the previous sites as above, but in addition the *next-to-last* site of the loop $L_i$ must avoid the *next* site $\rho(i+1)$ of the backbone (except of course for $i = |\rho|$, when this constraint is vacuous). For memories $\tau \geq 4$ the situation is more complicated, due to the presence of inter-loop restrictions, and we refrain from entering into details.

The above discussion can be summarized with identities, which are stated below for $\tau = 0$ and $\tau = 2$. To state the identities we first define the *generating functions with taboo set $A$*:

$$C_\tau^A(x, y; \beta) = \sum_{\omega \in \Omega_\tau(x,y): \omega \cap A = \emptyset} \beta^{|\omega|} \tag{2.15}$$

$$\widetilde{C}_\tau^{A,z}(x, y; \beta) = \sum_{\substack{\omega \in \Omega_\tau(x,y): \omega \cap A = \emptyset \\ \omega(|\omega|-1) \neq z}} \beta^{|\omega|}. \tag{2.16}$$

In both cases the sum is over walks which avoid the set of sites $A$; in the latter case we impose the additional restriction that the next-to-last site of $\omega$ is not $z$. Clearly, $\widetilde{C}_\tau^{A,z}(x, y; \beta) \leq C_\tau^A(x, y; \beta)$, and both quantities are decreasing functions of $\tau$ and of the set $A$. We also define $\rho[0, j)$ to be the set of sites $\{\rho(0), \ldots, \rho(j-1)\}$, for $j = 1, \ldots, |\omega|$, and let $\rho[0, 0)$ be the empty set. We can then write the *identities*

$$C_0(x, y; \beta) = \sum_{\rho \in \Omega(x,y)} \beta^{|\rho|} \prod_{j=0}^{|\rho|} C_0^{\rho[0,j)}(\rho(j), \rho(j); \beta) \tag{2.17}$$

$$C_2(x, y; \beta) = \sum_{\rho \in \Omega(x,y)} \beta^{|\rho|} \prod_{j=0}^{|\rho|} \widetilde{C}_2^{\rho[0,j),\rho(j+1)}(\rho(j), \rho(j); \beta), \tag{2.18}$$

where the $j = |\rho|$ term in this last equation should be interpreted as $C_2^{\rho[0,j)}$ (since the site $\rho(j+1)$ is nonexistent). For higher memories it is less straightforward to write the analogous identities, because of the inter-loop constraints; but by dropping those constraints we have immediately the *inequalities*

$$C_\tau(x, y; \beta) \leq \sum_{\rho \in \Omega(x,y)} \beta^{|\rho|} \prod_{j=0}^{|\rho|} C_\tau^{\rho[0,j)}(\rho(j), \rho(j); \beta) \qquad \text{for } \tau \geq 0 \tag{2.19}$$

$$C_\tau(x, y; \beta) \leq \sum_{\rho \in \Omega(x,y)} \beta^{|\rho|} \prod_{j=0}^{|\rho|} \widetilde{C}_\tau^{\rho[0,j),\rho(j+1)}(\rho(j), \rho(j); \beta) \qquad \text{for } \tau \geq 2. \tag{2.20}$$

## 2.3 Inequalities (first version)

It is certainly difficult to analyze the right sides of (2.17)–(2.20) exactly, but we wish to do something less ambitious: we will obtain upper bounds by relaxing the avoidance constraints on the attached loops. In particular, we can obtain upper bounds by replacing the restriction that the loop attached at $\rho(j)$ avoid all of $\rho[0,j)$ by the weaker restriction that it avoid only the smaller set

$$\rho[j-k,j) \equiv \{\rho(j-k),\ldots,\rho(j-1)\} \,, \tag{2.21}$$

where $k$ is a (small) fixed nonnegative integer. [For $k=0$, $\rho[j,j) = \varnothing$; and for $k > j$ we omit the nonexistent points $\rho(i)$ with $i < 0$.] Applying this to (2.17)/(2.19) leads to the inequality

$$C_\tau(x,y;\beta) \leq \sum_{\rho \in \Omega(x,y)} \beta^{|\rho|} \prod_{j=0}^{|\rho|} C_\tau^{\rho[j-k,j)}(\rho(j),\rho(j);\beta) \tag{2.22}$$

for any $\tau, k \geq 0$. Now the set of sites $\rho[j-k,j)$ is (for $j \geq k$) simply the range of a $(k-1)$-step self-avoiding walk starting at a nearest neighbour of $\rho(j)$; so we can get a further upper bound by taking the maximum over all such sets. Taking into account translation invariance, this leads us to define

$$\Lambda_\tau(k;\beta) \;=\; \max_A C_\tau^A(0,0;\beta) \,, \tag{2.23}$$

where the maximum ranges over all $k$-element sets $A$ which are the range of a $(k-1)$-step self-avoiding walk starting at a nearest neighbour of the origin. This maximum is obviously a nonincreasing function of $\tau$ and $k$. Clearly we have[1]

$$\begin{aligned} C_\tau(x,y;\beta) &\leq \sum_{\rho \in \Omega(x,y)} \beta^{|\rho|} \Lambda_\tau(0;\beta)\Lambda_\tau(1;\beta)\cdots\Lambda_\tau(k-1;\beta)\Lambda_\tau(k;\beta)^{|\rho|+1-k} \\ &\equiv \alpha_{\tau,k}(\beta)\, G(x,y;\beta\Lambda_\tau(k;\beta)) \,, \end{aligned} \tag{2.24}$$

where

$$\alpha_{\tau,k}(\beta) \;\equiv\; \left[\prod_{i=0}^{k-1} \Lambda_\tau(i;\beta)\right] \Lambda_\tau(k;\beta)^{1-k} \,. \tag{2.25}$$

Summing over $y \in \mathbb{Z}^d$, this gives

$$\frac{\chi_\tau(\beta)}{\alpha_{\tau,k}(\beta)} \;\leq\; \chi(\beta\Lambda_\tau(k;\beta)) \,. \tag{2.26}$$

This is our fundamental *loop-erasure-and-restoration inequality*, in its simplest form.

We shall argue below that

$$\lim_{\beta \uparrow \mu_\tau^{-1}} \frac{\chi_\tau(\beta)}{\alpha_{\tau,k}(\beta)} \;=\; \infty \tag{2.27}$$

for $\tau = 0, 2$, all $k \geq 0$ and all $d > 0$. We expect that (2.27) is true also for all $\tau \geq 4$, but we have not proved this, nor shall we make use of it. Given (2.27), it is now easy to obtain a lower bound on $\mu$, in the following way. By (2.27) and (2.26), $\chi_\tau(\mu_\tau^{-1}\Lambda_\tau(k;\mu_\tau^{-1})) = +\infty$. But we know from (2.5a) that $\chi(x) < \infty$ for $x < 1/\mu$. We thus conclude that

$$\frac{\Lambda_\tau(k;\mu_\tau^{-1})}{\mu_\tau} \;\geq\; \frac{1}{\mu} \,, \tag{2.28}$$

---

[1] The terms with $|\rho| < k-1$ here are being overcounted on the right-hand side, to simplify the form of the inequality.

or in other words

$$\mu \geq \frac{\mu_\tau}{\Lambda_\tau(k; \mu_\tau^{-1})} \quad \text{for } \tau = 0, 2, \ k \geq 0, \ d > 0 . \quad (2.29)$$

The bound (2.29) would also follow for $\tau \geq 4$, if (2.27) were proven for such $\tau$.

The claim (2.27) follows immediately from (2.7a) if $\alpha_{\tau,k}(\mu_\tau^{-1})$ is finite. And it follows easily from (2.25) that $\alpha_{\tau,k}(\mu_\tau^{-1})$ is finite precisely when $\Lambda_\tau(0; \mu_\tau^{-1}) = C_\tau(0, 0; \mu_\tau^{-1})$ is finite [recall that $\Lambda_\tau(i; \beta) \leq \Lambda_\tau(0; \beta)$ and $\Lambda_\tau(k; \beta) \geq 1$]. For $\tau = 0$ or 2 the finiteness of $C_\tau(0, 0; \mu_\tau^{-1})$ in dimensions $d > 2$ follows immediately from (2.9) and (2.14). This proves the claim (2.27) for $d > 2$. For $d \leq 2$, consider first $\tau = 0$. We have trivially that $\Lambda_0(k; \beta) \geq 1$. Also, it is well-known (see Section 3.1) that $\Lambda_0(1; \beta) \equiv C_0^{\{e\}}(0, 0; \beta)$ is uniformly bounded for $\beta \leq \frac{1}{2d}$, in all dimensions; hence the same is true for $\Lambda_0(i; \frac{1}{2d})$ for all $i \geq 1$. Finally, for $i = 0$ we have

$$\Lambda_0(0; \beta) = C_0(0, 0; \beta) \sim \begin{cases} \text{const} \times (1 - 2d\beta)^{-(2-d)/2} & \text{for } d < 2 \\ \text{const} \times \log(1 - 2d\beta) & \text{for } d = 2 \\ \text{const} & \text{for } d > 2 \end{cases} \quad (2.30)$$

as $\beta \uparrow \frac{1}{2d}$. Hence in any dimension $d > 0$ we see from (2.7a) and (2.30) that $\Lambda_0(0; \beta)$ diverges more slowly than $\chi_0(\beta)$ as $\beta \uparrow \frac{1}{2d}$; therefore, $\lim_{\beta \uparrow \frac{1}{2d}} \chi_0(\beta)/\alpha_{0,k}(\beta) = \infty$. The same argument can be used for $\tau = 2$, using also (2.13).

We remark that another possible approach to bounds on $\mu$ of this type, which we will not pursue further, would be to substitute some $\beta < \mu_\tau^{-1}$ in (2.26), rather than $\beta = \mu_\tau^{-1}$, and then to optimize over $\beta$. This surely gives an improvement when $d = 2$ and $k = 0$, but it requires some $\mu$-dependent *a priori* upper bound on $\chi(\beta)$, and the only available such bound is the very weak Hammersley-Welsh bound [19, 20] (for which explicit constants would be required).

To see explicitly what the bounds (2.29) entail, let us consider first the case $\tau = 0$, for which $\mu_0 = 2d$. Taking $k = 0$ and $k = 1$ gives $\Lambda_0(0; \frac{1}{2d}) = C_0(0, 0; \frac{1}{2d})$ and $\Lambda_0(1; \frac{1}{2d}) = C_0^{\{e\}}(0, 0; \frac{1}{2d})$ (by symmetry, where $e$ is any neighbour of the origin), so (2.29) becomes

$$\mu \geq \frac{2d}{C_0(0, 0; \frac{1}{2d})} \qquad [k = 0] \quad (2.31)$$

$$\mu \geq \frac{2d}{C_0^{\{e\}}(0, 0; \frac{1}{2d})} \qquad [k = 1] \quad (2.32)$$

For a discussion of the relation between (2.31) and some previously known results, see Section 5. The denominator on the right side of (2.31) is infinite in dimension $d \leq 2$, but this defect is remedied in (2.32): in Section 3.1 we shall prove the identity

$$C_0^{\{e\}}(0, 0; \frac{1}{2d}) = 2 - \frac{1}{C_0(0, 0; \frac{1}{2d})} \leq 2 \quad (2.33)$$

in all dimensions. This already improves the method used in [4], where an infinite denominator was always encountered for $d \leq 2$. In the trivial case $d = 1$, (2.32) gives the exact answer $\mu \geq 1$. For $d = 2$, (2.32) gives the rather poor bound $\mu \geq 2$. For $d = 3$, (2.32) already does better than all previously known bounds, yielding $\mu \geq 4.475817\ldots$ . In Section 3.1 we show how to carry out the computations at least in principle for arbitrary values of $k$, and we give explicit numerical results for $k = 0, 1, 2, 3, 4$ in dimensions $d = 2, 3, 4, 5, 6$. The resulting lower bounds on $\mu$ are tabulated in Table 2. In Section 6.3 we study the behaviour of these bounds as $d \to \infty$.

Next let us evaluate the bound (2.29) with memory $\tau = 2$. Here $\mu_2 = 2d - 1$. For $k = 0$, we conclude from (2.14) that

$$\mu \geq \frac{2d-1}{C_2(0,0;\frac{1}{2d-1})} = \frac{(2d-1)^2}{2d-2} \frac{1}{C_0(0,0;\frac{1}{2d})} . \tag{2.34}$$

This bound is nontrivial for $d > 2$, and is a factor

$$\frac{(2d-1)^2}{2d(2d-2)} = 1 + \frac{1}{2d(2d-2)} \tag{2.35}$$

better than the corresponding bound (2.31) based on $\tau = 0$. For $k = 1$ the situation is less simple: we will show in Section 3.2 that $C_2^{\{e\}}(0,0;\frac{1}{2d-1})$ can be obtained by solving a linear system of three equations in three unknowns. In Section 3.2 we show in fact how to carry out the computations at least in principle for arbitrary values of $k$, and we give explicit numerical results for $k = 0, 1, 2, 3, 4$ in dimensions $d = 2, 3, 4, 5, 6$. The resulting lower bounds on $\mu$ are tabulated in Table 2. In Section 6.3 we study the behaviour of these bounds as $d \to \infty$.

## 2.4 Inequalities (second version)

The foregoing inequalities are based on constraining the attached loops $L_i$ to avoid the *preceding* $k$ sites of the backbone $\rho$. For any memory $\tau \geq 2$, we can improve these results by using (2.18)/(2.20), i.e. by taking into account the further constraint that the next-to-last site of the loop avoid the *next* site of the backbone. The analysis given previously can be repeated almost verbatim in this case. We introduce

$$\widetilde{\Lambda}_\tau(k; \beta) = \max_{A,e} \widetilde{C}_\tau^{A,e}(0,0;\beta) , \tag{2.36}$$

where the maximum ranges over all $k$-element sets $A$ which are the range of a $(k-1)$-step self-avoiding walk starting at a nearest neighbour of the origin, and over all nearest neighbours $e$ of the origin satisfying $e \notin A$. The analogue of (2.24) becomes[2]

$$\begin{aligned}
C_\tau(x,y;\beta) &\leq \sum_{\rho \in \Omega(x,y)} \beta^{|\rho|} \Lambda_\tau(0;\beta) \cdots \Lambda_\tau(k-1;\beta) \widetilde{\Lambda}_\tau(k;\beta)^{|\rho|-k} \Lambda_\tau(k;\beta) \\
&\equiv \widetilde{\alpha}_{\tau,k}(\beta) G(x,y;\beta\widetilde{\Lambda}_\tau(k;\beta)),
\end{aligned} \tag{2.37}$$

where

$$\widetilde{\alpha}_{\tau,k}(\beta) = \left(\prod_{i=0}^{k} \Lambda_\tau(i;\beta)\right) \widetilde{\Lambda}_\tau(k;\beta)^{-k}. \tag{2.38}$$

---

[2] Here we are overcounting by neglecting the avoidance of the next backbone site, for the first $k$ sites of the backbone, to allow for a unified treatment of $|\rho| \leq k - 1$ and $|\rho| \geq k$.

Arguing as before, we have

$$\mu \geq \frac{\mu_\tau}{\widetilde{\Lambda}_\tau(k; \mu_\tau^{-1})} \qquad \text{for } \tau = 2, \ k \geq 0, \ d > 0 \tag{2.39}$$

(and again we expect, but have not proved, that this bound holds for all $\tau \geq 2$).

We will apply (2.39) with $\tau = 2$, which requires evaluation of $\widetilde{C}_2^{A,e}(0, 0; \mu_2^{-1})$. In Section 3.3 we will prove the identity

$$\widetilde{C}_2^{A,e}(0, 0; \beta) = \frac{1}{1 - \beta^2}[C_2^A(0, 0; \beta) - \beta C_2^A(0, e; \beta)], \tag{2.40}$$

from which the bound (2.39) can be computed once we know the values of $C_2^A(0, 0; \mu_2^{-1})$ and $C_2^A(0, e; \mu_2^{-1})$.

The use of (2.40) involves two improvements on our earlier method [4]. The first improvement is conceptual, in that the derivation here of (2.40) is simpler and avoids the combinatorial niceties encountered in [4]. The second is that our earlier method used only $|A| = 1$, and moreover constrained loops to avoid the previous backbone site only with their first and next-to-last sites, rather than entirely as in the methods of this paper.

Our method for numerically evaluating the denominators $\Lambda_0(k; \beta)$, $\Lambda_2(k; \beta)$ and $\widetilde{\Lambda}_2(k; \beta)$ which appear in the lower bounds on $\mu$ is described in Section 3. The numerical bounds resulting from (2.29) with $\tau = 0, 2$ and (2.39) with $\tau = 2$ are given in Table 2 as a function of the parameters $\tau$ and $k$. We have restricted attention to $\tau = 0$ and $\tau = 2$ due to our inability to compute the numerical values of $\Lambda_\tau(k; \beta)$ or $\widetilde{\Lambda}_\tau(k; \beta)$ for higher memories.

In Section 2.5 we will show how to improve on the bounds obtained so far.

## 2.5 Optimized bounds

Our method thus far has been based on taking the *maximum* over possible geometries for the incoming walk (i.e. the set $A$) in (2.23) or (2.36). This procedure is costly, because typically we expect that the incoming two steps of the backbone will be bent rather than straight, but the maximum for the small values of $|A|$ we are using corresponds to a straight backbone. In fact, given that there are on average about $\mu$ possible steps for a self-avoiding walk, the proportion of straight to bent steps should be roughly one to $\mu - 1$. (This is not exactly right, because the straight and bent steps have different probabilities for respecting self-avoidance; but it does indicate the expected order of magnitude.) In this section we show a way of partially taking this into account to obtain an improved bound. For concreteness, we consider the case of memory $\tau = 0$, with $k \equiv |A| = 2$.

Summing the identity (2.17) over $y \in \mathbb{Z}^d$, we have

$$\chi_0(\beta) = \sum_{\rho \in \Omega(0, \bullet)} \beta^{|\rho|} \prod_{j=0}^{|\rho|} C_0^{\rho[0,j)}(\rho(j), \rho(j); \beta) . \tag{2.41}$$

Relaxing the avoidance constraints on the attached loops, to $k = 2$, we obtain

$$\chi_0(\beta) \leq \sum_{\rho \in \Omega(0, \bullet)} \beta^{|\rho|} \prod_{j=0}^{|\rho|} C_0^{\rho[j-2,j)}(\rho(j), \rho(j); \beta) . \tag{2.42}$$

| $(\tau, k)$ | $d=2$ | $d=3$ | $d=4$ | $d=5$ | $d=6$ |
|---|---|---|---|---|---|
| (0,0) | 0 | 3.956 775 | 6.454 386 | 8.648 213 | 10.743 414 |
| (0,1) | 2 | 4.475 817 | 6.704 650 | 8.809 186 | 10.862 525 |
| (0,2) | 2.161 367 | 4.492 416 | 6.706 478 | 8.809 464 | 10.862 584 |
| (0,3) | 2.234 696 | 4.499 126 | 6.706 931 | 8.809 501 | 10.862 588 |
| (0,4) | 2.275 515 | 4.502 565 | 6.707 091 | 8.809 509 | 10.862 588 |
| (2,0) | 0 | 4.121 641 | 6.588 853 | 8.756 316 | 10.832 942 |
| (2,1) | 1.712 626 | 4.441 266 | 6.696 516 | 8.806 308 | 10.861 210 |
| (2,2) | 1.917 318 | 4.457 137 | 6.698 205 | 8.806 565 | 10.861 265 |
| (2,3) | 2.019 782 | 4.463 834 | 6.698 632 | 8.806 599 | 10.861 268 |
| (2,4) | 2.079 398 | 4.467 350 | 6.698 783 | 8.806 607 | 10.861 269 |
| $(\tilde{2},0)$ | 0 | 4.245 957 | 6.637 585 | 8.780 089 | 10.846 669 |
| $(\tilde{2},1)$ | 1.976 372 | 4.539 419 | 6.737 460 | 8.827 159 | 10.873 577 |
| $(\tilde{2},2)$ | 2.153 350 | 4.552 467 | 6.738 907 | 8.827 387 | 10.873 627 |
| $(\tilde{2},3)$ | 2.239 265 | 4.557 994 | 6.739 273 | 8.827 417 | 10.873 630 |
| $(\tilde{2},4)$ | 2.286 245 | 4.560 903 | 6.739 404 | 8.827 424 | 10.873 631 |
| $(0,2)_{opt}$ | 2.195 201 | 4.518 652 | 6.715 924 | 8.813 103 | 10.864 240 |
| $(0,3)_{opt}$ | 2.267 128 | 4.526 286 | 6.716 713 | 8.813 204 | 10.864 257 |
| $(0,4)_{opt}$ | 2.305 766 | 4.530 282 | 6.716 982 | 8.813 224 | 10.864 259 |
| $(2,2)_{opt}$ | 1.936 810 | 4.476 092 | 6.704 487 | 8.808 707 | 10.862 112 |
| $(2,3)_{opt}$ | 2.038 216 | 4.483 773 | 6.705 232 | 8.808 801 | 10.862 128 |
| $(2,4)_{opt}$ | 2.092 741 | 4.487 869 | 6.705 488 | 8.808 820 | 10.862 130 |
| $(\tilde{2},2)_{opt}$ | 2.165 878 | 4.562 269 | 6.742 085 | 8.828 430 | 10.874 022 |
| $(\tilde{2},3)_{opt}$ | 2.248 707 | 4.568 677 | 6.742 721 | 8.828 512 | 10.874 036 |
| $(\tilde{2},4)_{opt}$ | 2.290 302 | 4.572 140 | 6.742 945 | 8.828 529 | 10.874 038 |

Table 2: The lower bounds on $\mu$ of (2.29) for $\tau = 0, 2$ and of (2.39) for $\tau = \tilde{2}$, for $k = 1, 2, 3, 4$, and the optimized bounds using the method of Section 2.5, for dimensions $d = 2, 3, 4, 5, 6$. The numerical values have been truncated to give rigorous lower bounds. The lower line of the table provides the best lower bounds, except for $d = 2$. Some discussion of the table entries is given in Section 4.

Our previous approach was to replace the quantities $C_0^{\rho[j-2,j)}$ by the corresponding upper bounds based on using the *worst* 2-element set $A$ in place of $\rho[j-2,j)$ [cf. (2.23)/(2.24)]. Now let us try instead to distinguish the two possible cases,

$$\begin{aligned} A &= \text{straight (i.e. congruent to } \{e_1, 2e_1\}) \\ A &= \text{bent (i.e. congruent to } \{e_1, e_1 + e_2\}). \end{aligned}$$

Correspondingly, let us define

$$a \equiv C_0^{\{e_1, 2e_1\}}(0,0; \tfrac{1}{2d}) \tag{2.43a}$$
$$b \equiv C_0^{\{e_1, e_1+e_2\}}(0,0; \tfrac{1}{2d}). \tag{2.43b}$$

We only consider the case of

$$0 < b < a < 2d, \tag{2.44}$$

which is shown to be valid for our applications by numerical computation. It follows that[3]

$$\chi_0(\beta) \leq \sum_{\rho \in \Omega(0,\bullet)} \beta^{|\rho|} \Lambda_0(0;\beta) \Lambda_0(1;\beta) a^{\ell(\rho)} b^{|\rho|-1-\ell(\rho)}, \tag{2.45}$$

where $\ell$ denotes the number of straight vertices among $\rho(1), \ldots, \rho(n-1)$ [here $n \equiv |\rho|$]. Defining

$$c_n(\ell) \equiv \#\{\rho \in \Omega(0,\bullet) : |\rho| = n, \#(\text{straight vertices in } \rho) = \ell\}, \tag{2.46}$$

we can rewrite this as

$$\begin{aligned} \chi_0(\beta) &\leq \Lambda_0(0;\beta) \Lambda_0(1;\beta) \sum_{n=0}^{\infty} \sum_{\ell=0}^{n-1} c_n(\ell) \beta^n a^\ell b^{n-1-\ell} \\ &\equiv \Lambda_0(0;\beta) \Lambda_0(1;\beta) \sum_{n=0}^{\infty} \beta^n S(n;a,b) \end{aligned} \tag{2.47}$$

where

$$S(n;a,b) \equiv \sum_{\ell=0}^{n-1} c_n(\ell) a^\ell b^{n-1-\ell}. \tag{2.48}$$

As in the preceding subsections, we will let $\beta \uparrow \frac{1}{2d}$, and use the fact that

$$\lim_{\beta \uparrow \frac{1}{2d}} \frac{\chi_0(\beta)}{\Lambda_0(0;\beta) \Lambda_0(1;\beta)} = +\infty. \tag{2.49}$$

This together with (2.47) implies that

$$\limsup_{n \to \infty} S(n;a,b)^{1/n} \geq 2d. \tag{2.50}$$

---

[3] The terms with $|\rho| = 0, 1$ are here being overcounted on the right-hand side.

Therefore, if we can get upper bounds on $S(n; a, b)$ in terms of the $\{c_n\}$, then we will be able to deduce lower bounds on $\mu$.

**Example 1.** Since $a > b > 0$, we have trivially

$$S(n; a, b) \leq c_n a^{n-1} \tag{2.51}$$

and hence

$$\limsup_{n \to \infty} S(n; a, b)^{1/n} \leq \mu a . \tag{2.52}$$

Combining this with (2.50), we obtain $\mu \geq 2d/a$. This is, of course, our old bound (2.29) with $\tau = 0$, $k = 2$, based on using the *worst* set $A$.

**Example 2.** We claim that

$$c_n(\ell) \leq 2d \binom{n-1}{\ell} 1^\ell (2d-2)^{n-1-\ell} . \tag{2.53}$$

Indeed, there are $\binom{n-1}{\ell}$ ways of distributing $\ell$ straight vertices among the $n-1$ internal vertices of an $n$-step walk; and at each straight (resp. bent) vertex there are at most 1 (resp. $2d-2$) choices for the next step. The first step of the walk has, of course, $2d$ choices. [We remark that summing (2.53) over $\ell$ gives $c_n \leq 2d(2d-1)^{n-1}$, which is the trivial bound on $c_n$ in terms of memory-2 walks.] Inserting this bound into (2.48) and performing the sum, we conclude that

$$S(n; a, b) \leq 2d[a + (2d-2)b]^{n-1} \tag{2.54}$$

and hence

$$\limsup_{n \to \infty} S(n; a, b)^{1/n} \leq a + (2d-2)b . \tag{2.55}$$

Combining this with (2.50), we learn that

$$a + (2d-2)b \geq 2d \tag{2.56}$$

— an interesting fact, but one which unfortunately teaches us nothing about $\mu$.

Our approach now will be to *combine*, in an optimal way, these two ways of bounding $S(n; a, b)$. We shall use the first bound for $\ell \leq \lambda n$ and the second bound for $\ell > \lambda n$, with a suitable choice of $\lambda$. That is, we shall split $S = S_1 + S_2$, with

$$S_1(n; a, b) \equiv \sum_{\ell=0}^{\lfloor \lambda n \rfloor} c_n(\ell) a^\ell b^{n-1-\ell} \tag{2.57a}$$

$$S_2(n; a, b) \equiv \sum_{\ell=\lfloor \lambda n \rfloor + 1}^{n-1} c_n(\ell) a^\ell b^{n-1-\ell}. \tag{2.57b}$$

Since $a > b > 0$, we have trivially

$$S_1(n; a, b) \leq c_n a^{\lfloor \lambda n \rfloor} b^{n-1-\lfloor \lambda n \rfloor} \tag{2.58}$$

and hence
$$\limsup_{n\to\infty} S_1(n;a,b)^{1/n} \le \mu a^\lambda b^{1-\lambda} \ . \tag{2.59}$$

On the other hand, from (2.53) we have
$$S_2(n;a,b) \le 2d \sum_{\ell=\lfloor \lambda n \rfloor+1}^{n-1} \binom{n-1}{\ell} a^\ell [(2d-2)b]^{n-1-\ell} \ . \tag{2.60}$$

Using Stirling's formula, we find that the summand achieves its maximum at
$$\frac{\ell}{n} \approx \lambda_0 \equiv \frac{a}{a+(2d-2)b} \tag{2.61}$$

and is decreasing thereafter. So let us take $\lambda \in [\lambda_0, 1]$; then easy estimates show that
$$\limsup_{n\to\infty} S_2(n;a,b)^{1/n} \le \left(\frac{a}{\lambda}\right)^\lambda \left(\frac{(2d-2)b}{1-\lambda}\right)^{1-\lambda} \ . \tag{2.62}$$

Combining (2.50) with (2.59) and (2.62), we conclude that
$$\max\left[\mu a^\lambda b^{1-\lambda}, \left(\frac{a}{\lambda}\right)^\lambda \left(\frac{(2d-2)b}{1-\lambda}\right)^{1-\lambda}\right] \ge 2d \tag{2.63}$$

for all $\lambda \in [\lambda_0, 1]$. We will choose $\lambda \in [\lambda_0, 1]$ such that the second expression is just barely less than $2d$; then we can conclude that the first expression is $\ge 2d$. Now, the second expression equals $a + (2d-2)b \ge 2d$ [the inequality due to (2.56)] when $\lambda = \lambda_0$, and equals $a < 2d$ [the inequality due to our assumption (2.44)] when $\lambda = 1$, and is a continuous and strictly decreasing function of $\lambda$ in the interval $[\lambda_0, 1]$. Therefore, there is a unique $\lambda^* \in [\lambda_0, 1]$ such that
$$\left(\frac{a}{\lambda^*}\right)^{\lambda^*} \left(\frac{(2d-2)b}{1-\lambda^*}\right)^{1-\lambda^*} = 2d \ . \tag{2.64}$$

We then have the bound
$$\mu \ge \frac{2d}{a^{\lambda^*} b^{1-\lambda^*}} = \left(\frac{2d-2}{1-\lambda^*}\right)^{1-\lambda^*} \left(\frac{1}{\lambda^*}\right)^{\lambda^*} \ . \tag{2.65}$$

The above inequality can be used to improve on our previous bounds. For $|A|=2$, it can be applied directly. To apply it when $|A|>2$, we classify $A$ according to its first two steps, and distinguish the two possibilities by defining now [we write $A \equiv (a_1, a_2, a_3, ...)$]
$$\begin{aligned} a &\equiv \max_{A:\, a_1=e_1, a_2=2e_1} C_0^A(0,0;\tfrac{1}{2d}) \\ b &\equiv \max_{A:\, a_1=e_1, a_2=e_1+e_2} C_0^A(0,0;\tfrac{1}{2d}) \end{aligned} \tag{2.66a}$$

Then (2.45) still holds for these $a$ and $b$, and we can proceed in exactly the same way as for the $|A|=2$ case. The bounds for memory-2 and memory-$\tilde{2}$ can be optimized in the same way. The resulting bounds are given in Table 2.

# 3 Evaluation of loop generating functions

## 3.1 Simple random walk

In this section we shall reduce the computation of $\Lambda_0(k;\beta)$ to the evaluation of the simple-random-walk two-point function $C_0(0, x; \beta)$ at a suitable finite set of sites $x$. In particular, for computing the numerical bounds in Table 2 we need only the critical case $\beta = 1/2d$. Methods for obtaining the values of the critical two-point function are discussed in detail in Appendix B of [4], and a brief summary is given in Appendix A below. Of course the critical simple-random-walk two-point function is infinite in dimensions $d = 1, 2$; we briefly indicate the modifications needed to treat $d \leq 2$ at the end of this section.

We begin by deriving a recursion relation which relates the generating functions $C_0^A(y, x; \beta)$ and $C_0^{A \cup \{b\}}(y, x; \beta)$, where $b$ is a single site and $A$ is a finite set of sites which does not contain $b$. Applying inclusion-exclusion gives

$$C_0^{A \cup \{b\}}(y, x; \beta) = C_0^A(y, x; \beta) - \sum_{\substack{\omega \in \Omega_0(y, x) \\ \omega \ni b, \omega \cap A = \emptyset}} \beta^{|\omega|} . \tag{3.1}$$

Given a walk $\omega$ contributing to the sum on the right side, we break the walk into two pieces at its first visit to $b$. The generating function for walks which go from $y$ to $b$, which hit $b$ only once (namely at their last step), and which avoid the set $A$, is equal to $C_0^A(y, b; \beta) / C_0^A(b, b; \beta)$. Therefore

$$C_0^{A \cup \{b\}}(y, x; \beta) = C_0^A(y, x; \beta) - \frac{C_0^A(y, b; \beta) \, C_0^A(b, x; \beta)}{C_0^A(b, b; \beta)} . \tag{3.2}$$

When $A$ is the empty set, the right side can be computed in terms of the ordinary two-point function $C_0$, and then by iteration we can compute $C_0^A(y, x; \beta)$ for any finite set $A$.

An amusing special case is $y = x = 0$ and $A = \{e\}$, where $e$ is a nearest neighbour of the origin. Using the identity

$$C_0(0, 0; \beta) = 1 + 2d\beta C_0(0, e; \beta) \tag{3.3}$$

together with (3.2), we find

$$C_0^{\{e\}}(0, 0; \beta) = \left(1 - \frac{1}{(2d\beta)^2}\right) C_0(0, 0; \beta) + \frac{2}{(2d\beta)^2} - \frac{1}{(2d\beta)^2 C_0(0, 0; \beta)} . \tag{3.4}$$

This is finite as $\beta \uparrow \frac{1}{2d}$ in all dimensions $d > 0$ [since $C_0(0, 0; \beta)$ diverges as $\beta \uparrow \frac{1}{2d}$ more slowly than $(1 - 2d\beta)^{-1}$] and yields

$$C_0^{\{e\}}(0, 0; \tfrac{1}{2d}) = 2 - \frac{1}{C_0(0, 0; \tfrac{1}{2d})} \leq 2 . \tag{3.5}$$

In particular, we conclude that $\Lambda_0(i; \beta) \leq \Lambda_0(1; \beta) = C_0^{\{e\}}(0, 0; \beta) \leq 2$ for all $i \geq 1$ and all $\beta \leq \frac{1}{2d}$, in all dimensions.

The computation of $\Lambda_0(k;\beta)$ for any chosen $\beta$ is now reduced to a finite amount of labour. We simply list all the allowable sets $A$ of the given cardinality $k$, exploiting the obvious lattice symmetries, and then compute $C_0^A(0,0;\beta)$ for each such $A$ by iterating (3.2). For $|A|=1$, we have only one choice, $A=\{e\}$. For $|A|=2$, we compute the maximum over the two choices $A=\{e_1,2e_1\}$ and $A=\{e_1,e_1+e_2\}$, where $e_i$ are the canonical unit vectors. And so forth. Values for $C_0^A(0,0;\beta)$ are tabulated in Table 5 in the Appendix, and the resulting lower bounds on $\mu$ are given in Table 2.

In dimension $d>2$, we can perform these computations directly at the critical point $\beta=\frac{1}{2d}$: all quantities are finite. However, in dimension $d\leq 2$ the two-point function $C_0(0,x;\frac{1}{2d})$ is infinite for all $x$. Nevertheless we can show that $C_0^A(0,x;\frac{1}{2d})$ is finite whenever $A\neq\emptyset$, by working first at $\beta<\frac{1}{2d}$, and then letting $\beta\uparrow\frac{1}{2d}$. Note first that by the monotone convergence theorem we have

$$C_0^A(0,x;\tfrac{1}{2d}) = \lim_{\beta\uparrow\frac{1}{2d}} C_0^A(0,x;\beta). \tag{3.6}$$

Now the trick is to rewrite the case $A=\emptyset$ of (3.2), which is certainly valid for $\beta<\frac{1}{2d}$, in terms of a quantity that (unlike $C_0$) stays finite as $\beta\uparrow\frac{1}{2d}$. Such a quantity is the potential kernel $\Delta_0(x;\beta)\equiv C_0(0,0;\beta)-C_0(0,x;\beta)$ discussed in Section 2.1. Rewriting (3.2) with $A=\emptyset$ in terms of the potential kernel gives

$$C_0^{\{b\}}(y,x;\beta) = -\Delta_0(x-y;\beta)+\Delta_0(b-y;\beta)+\Delta_0(x-b;\beta)-\frac{\Delta_0(b-y;\beta)\,\Delta_0(x-b;\beta)}{C_0(0,0;\beta)}, \tag{3.7}$$

which in the limit $\beta\uparrow\frac{1}{2d}$ in dimension $d\leq 2$ reduces to the result of Proposition 11.6 of [15]:

$$C_0^{\{b\}}(y,x;\tfrac{1}{2d}) = -\Delta_0(x-y)+\Delta_0(b-y)+\Delta_0(x-b). \tag{3.8}$$

In particular this shows that $C_0^A(y,x;\frac{1}{2d})<\infty$ whenever $A\neq\emptyset$. Thus, once we have handled the step from $C_0$ to $C_0^{\{b\}}$ using (3.8), we can then advance to larger values of $|A|$ using (3.2) directly at $\beta=\frac{1}{2d}$, just as in higher dimensions. (An alternate method of computation is given in Theorem 14.11 of [15], but we have found the recursion easier to implement numerically.)

## 3.2 Memory-2 walk

The computation of our memory-2 lower bounds on $\mu$ has now been reduced to the evaluation of the generating functions $C_2^A(0,0;\mu_2^{-1})$ and $C_2^A(0,e;\mu_2^{-1})$ for finitely many finite sets $A$, just as was described in Section 3.1. The basic idea for the evaluation of these quantities is the same here as for simple random walk, but the recursion relation requires more care. Suppose that $b\notin A$. If $b=y$, we just have

$$C_2^{A\cup\{b\}}(y,x;\beta) = 0. \tag{3.9}$$

If $b\neq y$, we begin as for simple random walk by writing

$$C_2^{A\cup\{b\}}(y,x;\beta) = C_2(y,x;\beta) - \sum_{\substack{\omega\in\Omega_2(y,x)\\ \omega\ni b,\,\omega\cap A=\emptyset}} \beta^{|\omega|}. \tag{3.10}$$

To deal with the sum on the right side, we again want to cut the walk at the first time it hits $b$, but now the two pieces of the walk are no longer independent because of the memory-2 constraint. The sum on the right side is equal to

$$\sum_{\substack{\omega_1 \in \Omega_2(y,b), \omega_2 \in \Omega_2(b,x) \\ \omega_1, \omega_2 \cap A = \emptyset}} \beta^{|\omega_1|+|\omega_2|} \, \mathrm{I}[\omega_1(j) \neq b \text{ for } j < |\omega_1|] \, \mathrm{I}[\omega_1 \circ \omega_2 \in \Omega_2(y,x)]$$

$$= \sum_{f:|f|=1} \beta \sum_{\substack{\omega_1' \in \Omega_2(y,b+f) \\ \omega_1' \cap A = \emptyset}} \beta^{|\omega_1'|} \mathrm{I}[\omega_1' \not\ni b] \sum_{\substack{\omega_2 \in \Omega_2(b,x) \\ \omega_2 \cap A = \emptyset}} \beta^{|\omega_2|} \mathrm{I}[\omega_2(1) \neq b+f],$$

(3.11)

where $\omega_1 \circ \omega_2$ denotes the concatenation of the two walks. The sum over $\omega_1'$ exactly gives

$$C_2^{A \cup \{b\}}(y, b+f; \beta).$$

For the sum over $\omega_2$, we use repeated inclusion-exclusion to remove the constraint that $\omega_2(1) \neq b+f$, to obtain

$$\sum_{\substack{\omega_2 \in \Omega_2(b,x) \\ \omega_2 \cap A = \emptyset}} \beta^{|\omega_2|} \mathrm{I}[\omega_2(1) \neq b+f] = C_2^A(b,x;\beta) - \beta C_2^A(b+f,x;\beta)$$
$$+ \beta^2 C_2^A(b,x;\beta) - \beta^3 C_2^A(b+f,x;\beta) + \ldots$$

$$= \frac{1}{1-\beta^2} [C_2^A(b,x;\beta) - \beta C_2^A(b+f,x;\beta)].$$

(3.12)

The result is

$$C_2^{A \cup \{b\}}(y,x;\beta) = C_2^A(y,x;\beta)$$
$$- \frac{\beta}{1-\beta^2} \sum_{f:|f|=1} C_2^{A \cup \{b\}}(y, b+f; \beta) [C_2^A(b,x;\beta) - \beta C_2^A(b+f,x;\beta)].$$

(3.13)

The above unfortunately is not a closed-form expression for the left side, because of the occurrence of similar quantities on the right side. However, for a fixed $y$, (3.13) provides a systems of $2d$ linear equations for $2d$ unknowns, namely $\{C_2^{A \cup \{b\}}(y, b+f; \beta)\}_{|f|=1}$. Once we know these $2d$ quantities by solving the equations, then everything on the right hand side of (3.13) is known, and $C_2^{A \cup \{b\}}(y,x;\beta)$ can then be computed for general $x$. Fortunately, the number of unknowns is often reduced by symmetry. For example, for $y = 0$, $A = \emptyset$ and $b = e_1$, there are three inequivalent values of $b+f$, namely $0$, $2e_1$ and $e_1 + e_2$. Thus we can first solve the system of three equations in three unknowns which results by taking $x$ to be each of the three values $0$, $2e_1$ and $e_1 + e_2$, and once this system has been solved we can then compute the left side directly for any other value of $x$. Concretely, for $d = 3$, we get

$$C_2^{\{e_1\}}(0, 0; \tfrac{1}{5}) \approx 1.125805 \tag{3.14a}$$

$$C_2^{\{e_1\}}(0, 2e_1; \tfrac{1}{5}) \approx 0.074441 \tag{3.14b}$$

$$C_2^{\{e_1\}}(0, e_1+e_2; \tfrac{1}{5}) \approx 0.138449 \tag{3.14c}$$

For $|A| \geq 1$ we can proceed similarly; results are given in Table 6 in the Appendix.

The right side of (3.13) can be used directly at the critical point in more than two dimensions, but it contains infinities for $d = 2$ when $A$ is the empty set. In two dimensions a limiting argument is first used to deal with $A = \emptyset$, and then for $|A| \geq 1$ the recursion can be used directly as above. Surprisingly, the resulting bounds are worse than those obtained using simple random walk, and so we do not give the details of the limiting argument, but instead give only the result: for $b \neq y$,

$$\begin{aligned}C_2^{\{b\}}(y, x; \tfrac{1}{3}) &= \frac{1}{3} + \frac{2}{3}[\Delta_0(b-y) - \Delta_0(x-y)] + \Delta_0(x-b) \\ &\quad - \frac{1}{12} \sum_{f:|f|=1} C_2^{\{b\}}(y, b+f; \tfrac{1}{3}) \Delta_0(x-b-f).\end{aligned} \quad (3.15)$$

Results are discussed in Section 4.2.

### 3.3 Memory-2 loop with taboo on penultimate site

In this section, we derive the identity (2.40), which shows how the evaluation of $\widetilde{C}_2^{A,e}(0, 0; \beta)$ can be reduced to the evaluation of $C_2^A(0, 0; \beta)$ and $C_2^A(0, e; \beta)$. By definition,

$$\widetilde{C}_2^{A,e}(0, 0; \beta) = C_2^A(0, 0; \beta) - \sum_{\substack{\omega \in \Omega_2(0,0) \\ \omega \cap A = \emptyset, \omega(|\omega|-1) = e}} \beta^{|\omega|}. \quad (3.16)$$

By applying the inclusion-exclusion relation repeatedly as in (3.12), we obtain

$$\sum_{\substack{\omega \in \Omega_2(0,0) \\ \omega \cap A = \emptyset, \omega(|\omega|-1) = e}} \beta^{|\omega|} = \frac{\beta}{1-\beta^2}[C_2^A(0, e; \beta) - \beta C_2^A(0, 0; \beta)]. \quad (3.17)$$

Inserting this result into (3.16) gives

$$\widetilde{C}_2^{A,e}(0, 0; \beta) = \frac{1}{1-\beta^2}[C_2^A(0, 0; \beta) - \beta C_2^A(0, e; \beta)], \quad (3.18)$$

and hence the bound (2.39) can be computed once we know the values of $C_2^A(0, 0; \mu_2^{-1})$ and $C_2^A(0, e; \mu_2^{-1})$. For $|A| \geq 1$, (3.18) can be used directly at the critical point in all dimensions.

## 4 Discussion of results

### 4.1 Three or more dimensions

From Table 2 it can be seen that for $d \geq 3$ the $\tau = 2$ bound does better than $\tau = 0$ when $k = 0$, as was already shown at the end of Section 2.3. Perhaps surprisingly, for $k \geq 1$ this situation is reversed: higher memory means a more "sophisticated" bound, but it does not necessarily mean a *better* bound! In any case, the bounds with memory-$\tilde{2}$ do better than the corresponding memory-0 bounds.

In all cases with memory-0 or memory-2 in Table 2, the set $A$ giving the maximum of $C_0^A(0,0;\frac{1}{2d})$ is the straight line segment, as can be expected intuitively for small $|A|$. However for large $|A|$ in general it is not to be expected that the straight segment is optimal, and an example of an optimal $A$ which is not straight is given below for $d = 2$. (We thank Greg Lawler and Alain Sznitman for discussions on the non-optimality of straight $A$ for large $|A|$.) In Table 2 it is also the case that straight $A$ gives the maximum for the memory-$\tilde{2}$ bounds, although *a priori* there seems to be no compelling reason to expect this to be the case: in (3.18) straight $A$ will give the maximum individually for each of $C_2^A(0,0;\beta)$ and $C_2^A(0,e;\beta)$, but possibly not for the difference.

## 4.2 Two dimensions

Unfortunately, for $d = 2$ our methods do not produce good bounds on $\mu$, as can be seen from Table 2. For example, with $|A| = 4$ and $\tau = 0$ we have only $\mu \geq 2.275515$. To get a measure of the inherent limitation of the method in two dimensions, and in view of the relatively slow convergence of the bounds as $k$ increases, we computed the value of $2d/C_0^{L_k}(0,0;\frac{1}{2d})$ with $L_k$ the collinear set of $k$ sites joining $(-1,0)$ to $(-k,0)$. For large $k$ we do not expect that $C_0^A(0,0;\frac{1}{2d})$ is maximized over sets $A$ with $|A| = k$ by $C_0^{L_k}(0,0;\frac{1}{2d})$, and in fact already for $k = 5$ and $A = \{(-1,0),(-2,0),(-3,0),(-4,0),(-4,1)\}$ we have $C_0^A(0,0;\frac{1}{2d}) = 1.739044 > 1.738131 = C_0^{L_5}(0,0;\frac{1}{2d})$. Nevertheless $2d/C_0^{L_k}(0,0;\frac{1}{2d})$ does provide an *upper bound* on the best possible $\tau = (0,k)$ lower bound on $\mu$. For $k = 60$ we found $2d/C_0^{L_{60}}(0,0;\frac{1}{2d}) = 2.404210$, and by extrapolating the sequence for $k \leq 60$ to $k \to \infty$, we found a limiting value less than 2.42. Substantially better lower bounds on $\mu$ can be obtained by other methods [2, 5].

The situation for $\tau = 2$ is only slightly better. Solving the system (3.15) of three equations in three unknowns gives

$$C_2^{\{e_1\}}(0,0;\tfrac{1}{3}) \approx 1.751695 \tag{4.19a}$$

$$C_2^{\{e_1\}}(0,2e_1;\tfrac{1}{3}) \approx 0.649131 \tag{4.19b}$$

$$C_2^{\{e_1\}}(0,e_1+e_2;\tfrac{1}{3}) \approx 0.799587 \tag{4.19c}$$

This gives the very weak bound $\mu \geq 3/1.751695 \geq 1.712626$. By imposing the additional restriction that there should be no direct returns at the end of the loop [the memory-$\tilde{2}$ bound of inequality (2.39)], this is improved to $\mu \geq 1.976372$, which is comparable to the result obtained with memory-0.

Again for memory-2 and memory-$\tilde{2}$ we computed the bounds arising from $A = L_k$ for large $k$, to obtain an idea of the inherent limitation of the method. For memory-2 and $k = 58$ we found a bound of 2.268661, while for memory-$\tilde{2}$ and $k = 58$ we found a bound of 2.443124. Only marginally higher values result when these are extrapolated to $k \to \infty$. Since straight $A$ in general will not be optimal, the best possible bound on $\mu$ using memory-2 or memory-$\tilde{2}$ may in fact do worse than these values. In particular, for the memory-$\tilde{2}$ bounds straight $A$ already fails to provide a maximum when $k = 4$, where the optimal $A$ is the bent set $A = \{(-1, 0), (-2, 0), (-3, 0), (-3, 1)\}$.

Because of the poor results for $d = 2$ we have not done a rigorous error analysis of these $d = 2$ computations, but we do expect that they are correct to the given accuracy.

# 5 Comparison to the Ising model and the infrared bound

Our simplest (and worst) bound

$$\mu \geq \frac{2d}{C_0(0, 0; \frac{1}{2d})} \tag{5.1}$$

is in fact not new, as it is an immediate consequence of earlier results of Fisher [21] and Fröhlich, Simon and Spencer [22]. Also, this bound was proven by Lawler [23] for dimensions $d > 4$, via a different perspective on loop erasure.

Fisher's result is that the two-point spin correlation function $\langle \sigma_x \sigma_y \rangle_J$ of the nearest-neighbour ferromagnetic Ising model with inverse temperature $J$ (and zero external field) obeys the inequality

$$\langle \sigma_x \sigma_y \rangle_J \leq G(x, y; \tanh J), \tag{5.2}$$

from which he concluded that

$$\mu \geq \coth J_{c,Ising} , \tag{5.3}$$

where $J_{c,Ising}$ is the critical inverse temperature of the Ising model. The infrared bound of [22] gives

$$J_{c,Ising} \leq \frac{C_0(0, 0; \frac{1}{2d})}{2d}. \tag{5.4}$$

Combining these two inequalities yields

$$\mu \geq \coth\left(\frac{C_0(0, 0; \frac{1}{2d})}{2d}\right) . \tag{5.5}$$

This bound is an improvement on (5.1), and yields (5.1) when combined with the inequality $\coth x \geq x^{-1}$ (for positive $x$).

Table 3 gives the value of $\coth J_{c,Ising}$, using numerical estimates of $J_{c,Ising}$, and compares it with the bounds (5.1) and (5.5) as well as with the best rigorous lower bounds on $\mu$. Values of $J_{c,Ising}$ are taken from the exact solution $J_{c,Ising} = \frac{1}{2}\log(1 + \sqrt{2})$ for $d = 2$, from the Monte Carlo study [24] for $d = 3$, and from the series extrapolation results of [25] for $d = 4$ and [11] for $d = 5, 6$. If we accept the (nonrigorous for $d \geq 3$) numerical values of $J_{c,Ising}$, then our lower bounds are better than the best possible bound that can arise from (5.3) for $d \geq 4$, but not for $d = 2, 3$. However, for $d = 2$ the enumeration bound of Conway and Guttmann [5] does better than the Ising-model bound. The fact that our best bounds do better in high dimensions than (5.3) can be explained by the fact that our best bounds capture more terms of the $1/d$ expansion for $\mu$ than does the right side of (5.3); see Section 6.

|   | $d=2$ | $d=3$ | $d=4$ | $d=5$ | $d=6$ |
|---|---|---|---|---|---|
| $J_{c,Ising}$ | 0.440 687 | 0.221 659 5 (26) | 0.149 663 (34) | 0.113 917 (7) | 0.092 294 (7) |
| $2d/C_0(0,0;\frac{1}{2d})$ | 0 | 3.956 776 | 6.454 386 | 8.648 214 | 10.743 415 |
| $\coth[\frac{1}{2d}C_0(0,0;\frac{1}{2d})]$ | 1 | 4.040 663 | 6.505 948 | 8.686 723 | 10.774 424 |
| $\coth J_{c,Ising}$ | 2.414 213 | 4.585 07 (5) | 6.731 5 (15) | 8.816 2 (6) | 10.865 7 (9) |
| our best bound on $\mu$ | 2.305 766 | 4.572 140 | 6.742 945 | 8.828 529 | 10.874 038 |
| best bound on $\mu$ | 2.620 02 | 4.572 140 | 6.742 945 | 8.828 529 | 10.874 038 |

Table 3: Comparison of the Ising-model lower bound $\coth J_{c,Ising}$ with the best lower bounds on $\mu$. The values of $J_{c,Ising}$ are nonrigorous numerical estimates for $d = 3, 4, 5, 6$; errors in the last digit(s) are shown in parentheses. Other quantities are rounded to 6 digits, except for the lower bounds in the last two lines, which have been truncated.

## 6 High-$d$ behaviour of the lower bounds

The lace expansion has been used [12] to give a rigorous proof that as $d \to \infty$,

$$\mu = 2d - 1 - \frac{1}{2d} - \frac{3}{(2d)^2} - \frac{16}{(2d)^3} + O(d^{-4}) \ . \qquad (6.1)$$

The next term has been given in [26, 27],

$$\mu = 2d - 1 - \frac{1}{2d-1} - \frac{2}{(2d-1)^2} - \frac{11}{(2d-1)^3} - \frac{62}{(2d-1)^4} + \cdots \qquad (6.2a)$$

$$= 2d - 1 - \frac{1}{2d} - \frac{3}{(2d)^2} - \frac{16}{(2d)^3} - \frac{102}{(2d)^4} + \cdots , \qquad (6.2b)$$

but with no rigorous bound on the error. In this section we study the $1/d$ expansion for our (and other people's) lower bounds on $\mu$, and show that our best bound agrees with (6.1) up to and including the term of order $d^{-2}$.

In Section 6.1 we analyze briefly the $d$-dependence of some older lower bounds on $\mu$. It turns out that most of these methods have very poor behaviour as $d \to \infty$; this was, in fact, the original motivation for us to develop the loop-erasure-and-restoration method. In Section 6.2 we comment briefly on some results of Kesten [20], which are based on a precursor of our method. In Section 6.3 we analyze the high-$d$ behaviour of the loop-erasure-and-restoration bounds, and show that they capture the first few terms of the $1/d$ expansion of $\mu$. An interesting structure emerges from the comparison of how many terms are captured for different pairs $(\tau, k)$. In Section 6.4 we carry out an analogous analysis for the bounds based on comparison to the Ising model.

## 6.1 High-$d$ behaviour of the older lower-bound methods

We assume in this subsection that the reader is acquainted with the definition and fundamental properties of *bridges* and *irreducible bridges*; see e.g. [2, 13].

Let $b_n$ (resp. $i_n$) be the number of $n$-step bridges (resp. $n$-step irreducible bridges) starting at the origin and ending anywhere. By convention, $b_0 = 1$ and $i_0 = 0$. These counts satisfy the renewal equation

$$b_n = \delta_{n,0} + \sum_{k=1}^{n} i_k b_{n-k} . \tag{6.3}$$

We define for $\beta \geq 0$ the generating functions

$$B(\beta) = \sum_{n=0}^{\infty} b_n \beta^n \tag{6.4}$$

$$I(\beta) = \sum_{n=0}^{\infty} i_n \beta^n. \tag{6.5}$$

Now $0 \leq i_n \leq b_n \leq \mu^n$, and hence it follows from (6.3) that

$$B(\beta) = \frac{1}{1 - I(\beta)} \qquad \text{for } \beta \leq \mu^{-1}. \tag{6.6}$$

Also, it is known that

$$B(\beta) \begin{cases} \leq (1-\beta\mu)^{-1} < \infty & \text{for } 0 \leq \beta < \mu^{-1} \\ \to +\infty & \text{for } \beta \uparrow \mu^{-1} \\ = +\infty & \text{for } \beta > \mu^{-1} \end{cases} \tag{6.7}$$

$$I(\beta) \begin{cases} \leq \beta\mu < 1 & \text{for } 0 \leq \beta < \mu^{-1} \\ = 1 & \text{for } \beta = \mu^{-1} \\ = +\infty & \text{for } \beta > \mu^{-1}. \end{cases} \tag{6.8}$$

Now suppose that we have enumerated $i_1, \ldots, i_N$. (In practice one does this by enumerating $b_1, \ldots, b_N$ and solving (6.3) [2].) Then defining

$$I_N(\beta) \equiv \sum_{n=1}^{N} i_n \beta^n , \tag{6.9}$$

we have $I_N(\beta) \leq I(\beta) < 1$ for $\beta < \mu^{-1}$. It follows that

$$\mu \geq \mu_{*,N} \equiv \frac{1}{\beta_{*,N}} , \tag{6.10}$$

where $\beta_{*,N}$ is the unique positive solution of the polynomial equation $I_N(\beta) = 1$.

How does the bound (6.10) behave as $d \to \infty$ for fixed $N$? We are unable to give the precise asymptotic behaviour, but we can give an upper bound as follows: Trivially we have $i_n \leq b_n \leq (2d-1)^{n-1}$, because the first step of a bridge must be in the $+e_1$ direction, and for the remaining steps there are at most $2d - 1$ choices each. Therefore,

$$I_N(\beta) \leq \sum_{n=1}^{N}(2d-1)^{n-1}\beta^n . \tag{6.11}$$

In particular, the solution $\beta_{*,N}$ of $I_N(\beta_{*,N}) = 1$ must be greater than or equal to the solution $\beta_{**,N}$ of $\sum_{n=1}^{N}(2d-1)^{n-1}\beta_{**,N}^n = 1$. Now, it is easy to see that for each *fixed* $N$, as $d \to \infty$ we have

$$\beta_{**,N} = (2d-1)^{-(N-1)/N}\left[1 - O(d^{-1/N})\right] , \tag{6.12}$$

so that

$$\mu_{*,N} \equiv \frac{1}{\beta_{*,N}} \leq \frac{1}{\beta_{**,N}} = (2d-1)^{(N-1)/N}\left[1 + O(d^{-1/N})\right] . \tag{6.13}$$

So this method, for any fixed $N$, can never get even the correct leading asymptotic *order* (namely $\mu \sim d$) as $d \to \infty$.

In practice, this method gives the best currently available bounds for $d = 2$ [2, 5], but it is inferior to the loop-erasure-and-restoration method for $d \geq 3$.

Other early lower bounds on $\mu$ are those of Rennie [28],

$$\mu \geq 2d - \log d - 3 + \sqrt{2} + \log 2 - \sum_{j=3}^{\infty}\frac{\log(j+1)}{j[j - \log(j+1)]} , \tag{6.14}$$

and Hammersley [29],

$$\mu \geq 2d - \log(2d-1) - 1 . \tag{6.15}$$

These bounds get the correct first term in the large-$d$ expansion, but numerically they are rather poor.

Some other methods for proving lower bounds on $\mu$ in $d = 2, 3$ are given in [14]; but they do not appear to behave well as $d \to \infty$.

## 6.2 Kesten's bounds

A precursor of our method was used by Kesten [20] to prove that

$$\mu \geq \mu_{2r} - O(d^{-r}) \tag{6.16}$$

for all $r \geq 0$. Kesten's method involves loop erasure and restoration at the level of *counts* rather than generating functions. Since trivially $\mu \leq \mu_{2r}$, (6.16) implies that the $1/d$ expansion of $\mu_{2r}$ agrees with the first $r + 1$ terms of the $1/d$ expansion for $\mu$. Unfortunately, from the point of view of numerical estimates on $\mu$, (6.16) is not very helpful, since it is difficult to get good constants in the error term. Nor is it easy to compute the $1/d$ expansion of $\mu_{2r}$ for $r \geq 3$, so as to obtain the $1/d$ expansion for $\mu$ beyond three terms. However, using $r = 2$ in (6.16) together with (2.8), one obtains

$$\mu \geq 2d - 1 - \frac{1}{2d} + O(d^{-2}) , \tag{6.17}$$

which gives a bound agreeing with the first three terms of the $1/d$ expansion for $\mu$, albeit without good control of the error term.

## 6.3 $1/d$ expansion for the loop-erasure-and-restoration lower bounds

We now turn to the computation of the $1/d$ expansions for some of our lower bounds on $\mu$. We use the shorthand $s = \frac{1}{2d}$. The standard of comparison for all our bounds is the series (6.2),

$$\mu = s^{-1} - 1 - s - 3s^2 - 16s^3 - 102s^4 + \ldots , \qquad (6.18)$$

which is provably correct through order $s^3$ and presumably correct also at order $s^4$. As always, we classify our bounds according to the memory they use ($\tau = 0, 2$ or $\tilde{2}$) and how far back on the backbone they enforce avoidance of the attached loops ($k = 0, 1, 2, 3, \ldots$). We introduce the notation $\mu^{(\tau,k)}$ to denote the bound obtained by considering $(\tau, k)$-quantities.

Our most basic bound (2.31) is based on $\tau = 0$, $k = 0$; using (A.17), its $1/d$ expansion is

$$\mu^{(0,0)} = \frac{2d}{C_0(0,0; \frac{1}{2d})} = s^{-1} - 1 - 2s - 7s^2 - 35s^3 - 215s^4 + O(s^5) . \qquad (6.19)$$

This gets the first two terms correct, but misses the term of order $s$. The next simplest bound (2.32) is based on $\tau = 0$, $k = 1$; using (A.17) and (2.33), we obtain

$$\mu^{(0,1)} = \frac{2d}{C_0^{\{e\}}(0,0; \frac{1}{2d})} = s^{-1} - 1 - s - 4s^2 - 22s^3 - 143s^4 + O(s^5) . \qquad (6.20)$$

This gets the first three terms correct, and just barely misses the term of order $s^2$. As can be seen from Table 8 in the Appendix, taking $k = 2$ or $3$ does not improve on (6.20).

Next let us consider the bounds based on the memory-2 loop. The simplest such bound (2.34) is based on $\tau = 2$, $k = 0$; using (2.14) and (A.17), it becomes

$$\mu^{(2,0)} = \frac{2d - 1}{C_2(0,0; \frac{1}{2d-1})} = s^{-1} - 1 - s - 6s^2 - 35s^3 - 222s^4 + O(s^5). \qquad (6.21)$$

This gets the first three terms correct. The bound is further improved if we go to $k = 1$: making use of Table 9, we have

$$\mu^{(2,1)} = \frac{2d - 1}{C_2^{\{e\}}(0,0; \frac{1}{2d-1})} = s^{-1} - 1 - s - 4s^2 - 23s^3 - 150s^4 + O(s^5). \qquad (6.22)$$

The term of order $s^2$ is improved compared to (6.21), but the coefficient is still not correct. This bound based on $\tau = 2$, $k = 1$ is inferior (at order $s^3$) to the simpler bound based on $\tau = 0$, $k = 1$, as was the case for the numerical values of Table 2. Using Table 9, it can be seen that the expansion for $\mu^{(2,2)}$ is identical to (6.22).

Better memory-2 bounds can be obtained by insisting that the attached loops avoid also the next site on the backbone (Section 2.4). The simplest such bound is

$$\mu^{(\tilde{2},0)} = \frac{2d - 1}{\widetilde{C}_2^{\varnothing,e}(0,0; \frac{1}{2d-1})} = s^{-1} - 1 - s - 5s^2 - 29s^3 - 188s^4 + O(s^5), \qquad (6.23)$$

where we have used Table 10. This gets the first three terms correct, and does better at order $s^2$ than (6.21), but still has not yet got the correct coefficient $-3$. This coefficient is however captured correctly if we combine the constraints involving the previous and next steps on the backbone, i.e. if we go to $\tau = \tilde{2}$, $k = 1$. To see this, consider the loop generating functions $\widetilde{C}_2^{\{e_1\},-e_1}(0,0;\frac{1}{2d-1})$ and $\widetilde{C}_2^{\{e_1\},e_2}(0,0;\frac{1}{2d-1})$, which by symmetry are the only two geometries to be considered. Intuitively, the second should be greater than the first, since a walk which must avoid $e_1$ will prefer to return to the origin from $-e_1$, so the constraint is greater when this possibility is disallowed. In fact this intuition is borne out by the numerical results, and as can be seen from Table 10 it is also evident from the $1/d$ expansion. From Table 10, we have

$$\mu^{(\tilde{2},1)} = \frac{2d-1}{\widetilde{C}_2^{\{e_1\},e_2}(0,0;\frac{1}{2d-1})} = s^{-1} - 1 - s - 3s^2 - 18s^3 - 124s^4 + O(s^5), \tag{6.24}$$

which has the correct coefficient of order $s^2$, and as a bonus only misses the correct coefficient of order $s^3$ by 2. It is also clear from Table 10 that the expansion for $\mu^{(\tilde{2},2)}$ is identical to that of $\mu^{(\tilde{2},1)}$, through the order shown in (6.24).

The optimized bounds of Section 2.5 do not improve substantially on (6.24): we find that for the optimized bound using $k=2$ and $\tau = \tilde{2}$ the terms up to and including order $s^3$ are as in (6.24), while the coefficient of $s^4$ is improved slightly from $-124$ to $-122$.

In general we do not expect that there will be improvements below order $s^5$ when $k$ is increased beyond 1, for a given memory. For example, for $k=2$ the number of hexagons which pass through a specific next-nearest neighbour of the origin is only $O(d)$, and hence when multiplied by $\beta^6$ is an order $s^5$ effect. This can be viewed as a partial explanation of the small size of the improvements observed in the numerical bounds as $|A|$ increases beyond 1.

## 6.4 $1/d$ expansion for the Ising-model lower bounds

Let us first look at the bound (5.5) obtained by using comparison to the Ising model together with the infrared bound on $J_{c,Ising}$. This improves our basic $\tau = 0$, $k=0$ bound by virtue of the coth, and yields

$$\mu \geq s^{-1} - 1 - \tfrac{5}{3}s - \tfrac{20}{3}s^2 - \tfrac{1531}{45}s^3 + O(s^4). \tag{6.25}$$

We see that the coth makes only a slight improvement in the term of order $s$, and does *not* achieve the correct coefficient $-1$.

The method based on comparison to the Ising model is potentially more powerful than this, if one could get a better bound on $J_{c,Ising}$. The best possible result is obtained by inserting the exact (nonrigorous) expansion

$$J_{c,Ising} = \frac{1}{s^{-1} - 1 - \tfrac{4}{3}s - \tfrac{13}{3}s^2 - \tfrac{979}{45}s^3 - \tfrac{2009}{15}s^4 + \cdots} \tag{6.26}$$

derived by Fisher and Gaunt [26] into (5.3); the result is

$$\mu \geq \coth(J_{c,Ising}) = s^{-1} - 1 - s - 4s^2 - 21s^3 - \tfrac{394}{3}s^4 + \cdots. \tag{6.27}$$

This achieves the correct first three terms, but does not do as well as (6.24) on the term of order $s^2$. Of course, it is a highly nontrivial open problem to prove rigorously such good bounds on $J_{c,Ising}$.

# A  Simple random walk

## A.1  Numerical evaluation of the two-point function

To evaluate our lower bounds on $\mu$ numerically, it is necessary to know the numerical values of the critical simple-random-walk two-point function $C_0(0, x; \frac{1}{2d})$ for various values of $x$. An effective method of evaluating this quantity (as well as the subcritical two-point function) numerically to high precision, with rigorous error bounds, has been described in considerable detail in Appendix B of [4]. We now summarize briefly how the calculation goes in the critical case.

We begin with the integral formula

$$C_0(0, x; \tfrac{1}{2d}) = \int_{[-\pi,\pi]^d} \frac{d^d k}{(2\pi)^d} \frac{e^{ik \cdot x}}{1 - \widehat{D}(k)} \ . \tag{A.1}$$

where

$$\widehat{D}(k) = \frac{1}{d} \sum_{i=1}^{d} \cos k_i. \tag{A.2}$$

We then apply the identity

$$\frac{1}{A} = \int_0^\infty e^{-At} dt \ , \tag{A.3}$$

which is valid for $A > 0$, with $A = 1 - \widehat{D}(k)$. This factorizes the integrals over $k_1, \ldots, k_d$, which can then be performed to give

$$C_0(0, x; \tfrac{1}{2d}) = d \int_0^\infty \prod_{i=1}^{d} f_{|x_i|}(t) \, dt \ , \tag{A.4}$$

where

$$f_N(z) \equiv e^{-z} I_N(z) \equiv \frac{1}{2\pi} \int_{-\pi}^{\pi} e^{-z(1-\cos\theta)} \cos N\theta \, d\theta \tag{A.5}$$

and $I_N(z)$ is the modified Bessel function. This transforms the $d$-dimensional integral (A.1) into a one-dimensional integral over a semi-infinite interval.

The integrand in (A.4) decays as $t \to \infty$ only as a power. We speed up the decay by making the change of variables $t = e^u$, obtaining

$$C_0(0, x; \tfrac{1}{2d}) = d \int_{-\infty}^{\infty} F(u) \, du \tag{A.6}$$

| $x$ | $\Delta_0(x)$ | $C_0(0, x; \frac{1}{2d})$ | | | |
|---|---|---|---|---|---|
| | $d = 2$ | $d = 3$ | $d = 4$ | $d = 5$ | $d = 6$ |
| (0) | 0 | 1.5163860744 | 1.2394671218 | 1.1563081248 | 1.1169633732 |
| (1) | 1 | 0.5163860744 | 0.2394671218 | 0.1563081248 | 0.1169633732 |
| (1,1) | 1.2732395447 | 0.3311486174 | 0.1017176302 | 0.0474085960 | 0.0271774706 |
| (1,1,1) | — | 0.2614701416 | 0.0618723811 | 0.0222517907 | 0.0100651251 |
| (1,1,1,1) | — | — | 0.0447274307 | 0.0133523237 | 0.0049990827 |
| (1,1,1,1,1) | — | — | — | 0.0092253734 | 0.0029872751 |
| (2) | 1.4535209105 | 0.2573359025 | 0.0659640719 | 0.0275043553 | 0.0148223998 |
| (2,1) | 1.5464790895 | 0.2155896361 | 0.0436586366 | 0.0139794831 | 0.0058409498 |
| (2,1,1) | — | 0.1917916659 | 0.0334570990 | 0.0089609415 | 0.0030848645 |
| (2,1,1,1) | — | — | 0.0275824802 | 0.0065163319 | 0.0019448478 |
| (2,2) | 1.6976527263 | 0.1683310508 | 0.0259898362 | 0.0062387819 | 0.0019599801 |
| (2,2,1) | — | 0.1569524280 | 0.0221867673 | 0.0047601264 | 0.0013047592 |
| (3) | 1.7211254632 | 0.1652707962 | 0.0262936339 | 0.0068995628 | 0.0024959268 |
| (3,1) | 1.7615031763 | 0.1531389140 | 0.0217691587 | 0.0048774493 | 0.0014526312 |
| (3,1,1) | — | 0.1441957255 | 0.0189286425 | 0.0038130772 | 0.0009927439 |
| (3,2) | 1.8488263632 | 0.1324510884 | 0.0159271735 | 0.0029340473 | 0.0006998942 |
| (4) | 1.9079745896 | 0.1217332189 | 0.0137700477 | 0.0024716782 | 0.0006024105 |
| (4,1) | 1.9295817894 | 0.1171305125 | 0.0125592552 | 0.0020829365 | 0.0004528519 |
| (5) | 2.0516093163 | 0.0966064672 | 0.0085112166 | 0.0011537265 | 0.0002044795 |

Table 4: Numerical values of $\Delta_0(x)$ for $d = 2$, and of $C_0(0, x; \frac{1}{2d})$ for $d = 3, 4, 5, 6$, for the values of $x$ needed to compute the lower bounds on $\mu$. The values are rounded to ten digits after the decimal point. Final components of $x$ which are not shown are equal to zero.

where
$$F(u) = e^u \prod_{i=1}^{d} f_{|x_i|}(e^u). \quad (A.7)$$

We then use standard methods to bound the difference between the integral (A.6) and an infinite Riemann sum, truncate the Riemann sum with bounds on the omitted tails, and evaluate the resulting finite Riemann sum, using the large-$z$ asymptotic expansion for the modified Bessel function to deal with large $t$ and a truncated Taylor series for the modified Bessel function to deal with the remaining $t$. We also take into account possible round-off errors in the numerical computations. The result is that we obtain the values in Table 4. The values of $\Delta_0(x)$ for $d = 2$, which are known exactly, have been computed using the algorithm described in Section 15 of [15].

Table 5 gives numerical values of $C_0^A(0, 0; \frac{1}{2d})$, computed as described in Section 3.1. Table 6 gives the values of $C_2^A(0, 0; \frac{1}{2d-1})$, computed using the method of Section 3.2. Table 7 gives numerical values for $\tilde{C}_2^{A,e}(0, 0; \frac{1}{2d-1})$, computed using the method of Section 3.3.

| $A$ | $d=2$ | $d=3$ | $d=4$ | $d=5$ | $d=6$ |
|---|---|---|---|---|---|
| $\varnothing$ | $\infty$ | 1.516386 | 1.239467 | 1.156308 | 1.116963 |
| $\{e_1\}$ | 2 | 1.340537 | 1.193202 | 1.135179 | 1.104715 |
| $\{e_1, 2e_1\}$ | 1.850680 | 1.335584 | 1.192876 | 1.135143 | 1.104709 |
| $\{e_1, e_1+e_2\}$ | 1.735910 | 1.322546 | 1.190625 | 1.134570 | 1.104514 |
| $\{e_1, 2e_1, 3e_1\}$ | 1.789955 | 1.333592 | 1.192796 | 1.135138 | 1.104709 |
| $\{e_1, 2e_1, 2e_1+e_2\}$ | 1.767994 | 1.330950 | 1.192528 | 1.135103 | 1.104703 |
| $\{e_1, e_1+e_2, 2e_1+e_2\}$ | 1.695733 | 1.320170 | 1.190465 | 1.134555 | 1.104512 |
| $\{e_1, e_1+e_2, e_1+2e_2\}$ | 1.658351 | 1.317957 | 1.190277 | 1.134530 | 1.104507 |
| $\{e_1, e_1+e_2, e_1+e_2+e_3\}$ | — | 1.314688 | 1.189764 | 1.134435 | 1.104485 |
| $\{e_1, e_1+e_2, e_2\}$ | 1.357421 | 1.226256 | 1.153726 | 1.115672 | 1.093040 |

Table 5: Numerical values of $C_0^A(0, 0; \frac{1}{2d})$ for $d = 2, 3, 4, 5, 6$, for various choices of $A$. The values are rounded to 6 digits after the decimal point.

## A.2  $1/d$ expansions

This section contains $1/d$ expansions for several relevant quantities. As before, we use the shorthand $s = \frac{1}{2d}$. First, the following are obtained by directly integrating powers of cosines.

$$\int_{[-\pi,\pi]^d} \frac{d^d k}{(2\pi)^d} \hat{D}(k)^n = \begin{cases} s & (n=2) \\ 3s^2 - 3s^3 & (n=4) \\ 15s^3 - 45s^4 + 40s^5 & (n=6) \\ 105s^4 - 630s^5 + 1435s^6 - 1155s^7 & (n=8) \\ 945s^5 + O(s^6) & (n=10) \\ O(s^6) & (n \geq 12) \end{cases} \quad (A.8)$$

$$\int_{[-\pi,\pi]^d} \frac{d^d k}{(2\pi)^d} \hat{D}(k)^n \cos 2k_1 = \begin{cases} s^2 & (n=2) \\ 6s^3 - 8s^4 & (n=4) \\ 45s^4 - 165s^5 + 165s^6 & (n=6) \\ 420s^5 + O(s^6) & (n=8) \\ O(s^6) & (n \geq 10) \end{cases} \quad (A.9)$$

$$\int_{[-\pi,\pi]^d} \frac{d^d k}{(2\pi)^d} \hat{D}(k)^n \cos k_1 \cos k_2 = \begin{cases} 2s^2 & (n=2) \\ 12s^3 - 24s^4 & (n=4) \\ 90s^4 - 450s^5 + 660s^6 & (n=6) \\ 840s^5 + O(s^6) & (n=8) \\ O(s^6) & (n \geq 10) \end{cases} \quad (A.10)$$

| $A$ | $d=2$ | $d=3$ | $d=4$ | $d=5$ | $d=6$ |
|---|---|---|---|---|---|
| $\varnothing$ | $\infty$ | 1.213109 | 1.062400 | 1.027829 | 1.015421 |
| $\{e_1\}$ | 1.751695 | 1.125805 | 1.045320 | 1.021995 | 1.012778 |
| $\{e_1, 2e_1\}$ | 1.564686 | 1.121796 | 1.045056 | 1.021965 | 1.012773 |
| $\{e_1, e_1+e_2\}$ | 1.481256 | 1.113713 | 1.043739 | 1.021661 | 1.012681 |
| $\{e_1, 2e_1, 3e_1\}$ | 1.485308 | 1.120113 | 1.044989 | 1.021961 | 1.012773 |
| $\{e_1, 2e_1, 2e_1+e_2\}$ | 1.465260 | 1.118079 | 1.044790 | 1.021935 | 1.012769 |
| $\{e_1, e_1+e_2, 2e_1+e_2\}$ | 1.423896 | 1.111657 | 1.043606 | 1.021649 | 1.012680 |
| $\{e_1, e_1+e_2, e_1+2e_2\}$ | 1.388210 | 1.110045 | 1.043471 | 1.021631 | 1.012677 |
| $\{e_1, e_1+e_2, e_1+e_2+e_3\}$ | — | 1.107650 | 1.043096 | 1.021561 | 1.012660 |
| $\{e_1, e_1+e_2, e_2\}$ | 1.211771 | 1.071615 | 1.031850 | 1.017013 | 1.010433 |

Table 6: Numerical values of $C_2^A(0,0;\frac{1}{2d-1})$ for $d=2,3,4,5,6$, for various choices of $A$. The values are rounded to 6 digits after the decimal point.

| $A, e$ | $d=2$ | $d=3$ | $d=4$ | $d=5$ | $d=6$ |
|---|---|---|---|---|---|
| $\varnothing, e$ | $\infty$ | 1.177591 | 1.054600 | 1.025047 | 1.014136 |
| $\{e_1\}, -e_1$ | 1.304467 | 1.097370 | 1.038115 | 1.019313 | 1.011519 |
| $\{e_1\}, e_2$ | 1.349274 | 1.101462 | 1.038967 | 1.019581 | 1.011627 |
| $\{e_1, 2e_1\}, -e_1$ | 1.193791 | 1.093963 | 1.037871 | 1.019285 | 1.011514 |
| $\{e_1, 2e_1\}, e_2$ | 1.238380 | 1.098305 | 1.038744 | 1.019554 | 1.011622 |
| $\{e_1, e_1+e_2\}, -e_1$ | 1.140666 | 1.086889 | 1.036639 | 1.018992 | 1.011424 |
| $\{e_1, e_1+e_2\}, e_2$ | 1.210882 | 1.094439 | 1.038093 | 1.019408 | 1.011579 |
| $\{e_1, e_1+e_2\}, -e_2$ | 1.170686 | 1.090659 | 1.037469 | 1.019256 | 1.011531 |
| $\{e_1, e_1+e_2\}, e_3$ | — | 1.091432 | 1.037558 | 1.019272 | 1.011535 |

Table 7: Numerical values of $\widetilde{C}_2^{A,e}(0,0;\frac{1}{2d-1})$ for $d=2,3,4,5,6$, for various choices of $A, e$. The values are rounded to 6 digits after the decimal point.

$$\int_{[-\pi,\pi]^d} \frac{d^d k}{(2\pi)^d} \hat{D}(k)^n \cos 3k_1 = \begin{cases} s^3 & (n=3) \\ 10s^4 - 15s^5 & (n=5) \\ 105s^5 + O(s^6) & (n=7) \\ O(s^6) & (n \geq 9) \end{cases} \quad (A.11)$$

$$\int_{[-\pi,\pi]^d} \frac{d^d k}{(2\pi)^d} \hat{D}(k)^n \cos 2k_1 \cos k_2 = \begin{cases} 3s^3 & (n=3) \\ 30s^4 - 70s^5 & (n=5) \\ 315s^5 + O(s^6) & (n=7) \\ O(s^6) & (n \geq 9) \end{cases} \quad (A.12)$$

$$\int_{[-\pi,\pi]^d} \frac{d^d k}{(2\pi)^d} \hat{D}(k)^n \cos k_1 \cos k_2 \cos k_3 = \begin{cases} 6s^3 & (n=3) \\ 60s^4 - 180s^5 & (n=5) \\ 630s^5 + O(s^6) & (n=7) \\ O(s^6) & (n \geq 9) \end{cases} \quad (A.13)$$

The above, combined with the expansion

$$\frac{1}{1-\hat{D}} = \sum_{m=0}^{n} \hat{D}^m + \frac{\hat{D}^{n+1}}{1-\hat{D}} \quad (n \geq 0), \quad (A.14)$$

and estimates on errors using Lemma B.1 of [30], i.e.

$$\int_{[-\pi,\pi]^d} \frac{d^d k}{(2\pi)^d} \frac{|\hat{D}(k)|^m}{[1-\hat{D}(k)]^n} = O(s^{m/2}), \quad (d > 2n, \, m \geq 0) \quad (A.15)$$

leads to expansions for the integrals

$$I_{n,m}(x) = \int_{[-\pi,\pi]^d} \frac{d^d k}{(2\pi)^d} \frac{\hat{D}(k)^m e^{ik \cdot x}}{[1-\hat{D}(k)]^n}. \quad (A.16)$$

In particular for $C_0(0, x; \frac{1}{2d}) = I_{1,0}(x)$, we have

$$\begin{align} I_{10}(0) &= 1 + s + 3s^2 + 12s^3 + 60s^4 + 355s^5 + O(s^6) & (A.17) \\ I_{10}(e_1) &= s + 3s^2 + 12s^3 + 60s^4 + 355s^5 + O(s^6) & (A.18) \\ I_{10}(2e_1) &= s^2 + 6s^3 + 37s^4 + 255s^5 + O(s^6) & (A.19) \\ I_{10}(e_1 + e_2) &= 2s^2 + 12s^3 + 66s^4 + 390s^5 + O(s^6) & (A.20) \\ I_{10}(3e_1) &= s^3 + 10s^4 + 90s^5 + O(s^6) & (A.21) \\ I_{10}(2e_1 + e_2) &= 3s^3 + 30s^4 + 245s^5 + O(s^6) & (A.22) \\ I_{10}(e_1 + e_2 + e_3) &= 6s^3 + 60s^4 + 450s^5 + O(s^6). & (A.23) \end{align}$$

It can also be shown that

$$\sup_{|x|>3} I_{10}(x) = O(s^4). \quad (A.24)$$

Analogous expansions for higher $I_{n,m}$ will be given and used in [12].

| $A$ | $C_0^A(0,0;\frac{1}{2d})$ |
|---|---|
| $\varnothing$ | $1 + s + 3s^2 + 12s^3 + 60s^4 + 355s^5 + O(s^6)$ |
| $\{e_1\}$ | $1 + s + 2s^2 + 7s^3 + 35s^4 + 215s^5 + O(s^6)$ |
| $\{e_1, 2e_1\}$ | $1 + s + 2s^2 + 7s^3 + 35s^4 + 215s^5 + O(s^6)$ |
| $\{e_1, e_1 + e_2\}$ | $1 + s + 2s^2 + 7s^3 + 34s^4 + 202s^5 + O(s^6)$ |
| $\{e_1, 2e_1, 3e_1\}$ | $1 + s + 2s^2 + 7s^3 + 35s^4 + 215s^5 + O(s^6)$ |
| $\{e_1, 2e_1, 2e_1 + e_2\}$ | $1 + s + 2s^2 + 7s^3 + 35s^4 + 215s^5 + O(s^6)$ |
| $\{e_1, e_1 + e_2, 2e_1 + e_2\}$ | $1 + s + 2s^2 + 7s^3 + 34s^4 + 202s^5 + O(s^6)$ |
| $\{e_1, e_1 + e_2, e_1 + 2e_2\}$ | $1 + s + 2s^2 + 7s^3 + 34s^4 + 202s^5 + O(s^6)$ |
| $\{e_1, e_1 + e_2, e_1 + e_2 + e_3\}$ | $1 + s + 2s^2 + 7s^3 + 34s^4 + 202s^5 + O(s^6)$ |
| $\{e_1, e_1 + e_2, e_2\}$ | $1 + s + s^2 + 2s^3 + 14s^4 + 115s^5 + O(s^6)$ |

Table 8: $1/d$ expansions for $C_0^A(0,0;\frac{1}{2d})$ for various choices of $A$.

Finally, we turn to the $1/d$ expansions for loop-generating functions with taboo set. Beginning with the $1/d$ expansions for $C_0(y, x; \frac{1}{2d}) = I_{10}(x - y)$ given above, and then using the recursion (3.1), we obtain the $1/d$ expansions for $C_0^A(0, 0; \frac{1}{2d})$ given in Table 8. The $1/d$ expansions for memory-2 quantities are more difficult to handle. As described under (3.13), we first solve (3.13) for $\{C_2^{A \cup \{b\}}(y, b + f)\}_{|f|=1}$, for fixed $y$. At this stage, special care is needed to make full use of symmetry, since naively (3.13) is a system of equations for $2d$ unknowns (and here $d \nearrow \infty$). By symmetry, we can reduce (3.13) to a system of equations for a number of unknowns which is uniformly bounded in $d$ (at least for small $|A|$ and $|y|, |x|$), and obtain the results in Table 9. Then, using (3.18), we compute the $1/d$ expansions for $\tilde{C}_2^{A,e}(0, 0)$ given in Table 10.

# Acknowledgements

We wish to thank Tony Guttmann for helpful comments and for sending us his unpublished data.

The authors' research has been supported in part by Natural Sciences and Engineering Research Council of Canada grant A9351 (G.S.) and U.S. National Science Foundation grants DMS-8911273 and DMS-9200719 (A.D.S). Acknowledgment is also made to the donors of the Petroleum Research Fund, administered by the American Chemical Society, for partial support of this research under grants 21091–AC7 and 25553–AC7B–C (A.D.S.). The work of G.S. was carried out during visits to the Centre de Physique Théorique of the Ecole Polytechnique at Palaiseau and the Forschungsinstitut für Mathematik of the ETH at Zürich, and he thanks these institutions and their members for their kind hospitality. Numerical calculations were done using FORTRAN and Mathematica on an IBM RS/6000–350 of the Kitahara group of the Department of Applied Physics, Tokyo Institute of Technology.

| $A$ | $C_2^A(0,0;\frac{1}{2d-1})$ |
|---|---|
| $\varnothing$ | $1 + s^2 + 7s^3 + 43s^4 + 278s^5 + O(s^6)$ |
| $\{e_1\}$ | $1 + s^2 + 5s^3 + 29s^4 + 188s^5 + O(s^6)$ |
| $\{e_1, 2e_1\}$ | $1 + s^2 + 5s^3 + 29s^4 + 188s^5 + O(s^6)$ |
| $\{e_1, e_1+e_2\}$ | $1 + s^2 + 5s^3 + 29s^4 + 182s^5 + O(s^6)$ |

Table 9: $1/d$ expansions for $C_2^A(0,0;\frac{1}{2d-1})$ for various choices of $A$.

| $A, e$ | $\widetilde{C}_2^{A,e}(0,0;\frac{1}{2d-1})$ |
|---|---|
| $\varnothing, e$ | $1 + s^2 + 6s^3 + 36s^4 + 235s^5 + O(s^6)$ |
| $\{e_1\}, -e_1$ | $1 + s^2 + 4s^3 + 22s^4 + 146s^5 + O(s^6)$ |
| $\{e_1\}, e_2$ | $1 + s^2 + 4s^3 + 23s^4 + 154s^5 + O(s^6)$ |
| $\{e_1, 2e_1\}, -e_1$ | $1 + s^2 + 4s^3 + 22s^4 + 146s^5 + O(s^6)$ |
| $\{e_1, 2e_1\}, e_2$ | $1 + s^2 + 4s^3 + 23s^4 + 154s^5 + O(s^6)$ |
| $\{e_1, e_1+e_2\}, -e_1$ | $1 + s^2 + 4s^3 + 22s^4 + 140s^5 + O(s^6)$ |
| $\{e_1, e_1+e_2\}, e_2$ | $1 + s^2 + 4s^3 + 23s^4 + 152s^5 + O(s^6)$ |
| $\{e_1, e_1+e_2\}, -e_2$ | $1 + s^2 + 4s^3 + 23s^4 + 148s^5 + O(s^6)$ |
| $\{e_1, e_1+e_2\}, e_3$ | $1 + s^2 + 4s^3 + 23s^4 + 148s^5 + O(s^6)$ |

Table 10: $1/d$ expansions for $\widetilde{C}_2^{A,e}(0,0;\frac{1}{2d-1})$ for various choices of $A, e$.